\documentclass[aps,floatfix,nofootinbib,amsmath,amssymb,amsfonts,notitlepage,superscriptaddress, twocolumn]{revtex4-1}

\usepackage[T1]{fontenc}
\usepackage[latin9]{inputenc}
\usepackage{lmodern} % load a font with all the characters
\usepackage{color}
\usepackage{bbold}
\usepackage{dsfont}
\usepackage{graphicx}
\usepackage[caption=false]{subfig}
\usepackage[colorlinks]{hyperref}
\usepackage[bottom]{footmisc}
\usepackage{enumerate}
\usepackage{float}
\usepackage{mathtools}

\usepackage{empheq}
\usepackage{tikz}
\usetikzlibrary{patterns,tikzmark, matrix,decorations.pathreplacing, calc, positioning,fit}

\usepackage{hf-tikz}

\usepackage{fancyhdr}

\usepackage[normalem]{ulem}
\DeclareMathAlphabet\mbc{OMS}{cmsy}{b}{n}
\begin{document}

\global\long\def\eqn#1{\begin{align}#1\end{align}}
\global\long\def\vec#1{\overrightarrow{#1}}
\global\long\def\ket#1{\left|#1\right\rangle }
\global\long\def\bra#1{\left\langle #1\right|}
\global\long\def\bkt#1{\left(#1\right)}
\global\long\def\sbkt#1{\left[#1\right]}
\global\long\def\cbkt#1{\left\{#1\right\}}
\global\long\def\abs#1{\left\vert#1\right\vert}
\global\long\def\cev#1{\overleftarrow{#1}}
\global\long\def\der#1#2{\frac{{d}#1}{{d}#2}}
\global\long\def\pard#1#2{\frac{{\partial}#1}{{\partial}#2}}
\global\long\def\re{\mathrm{Re}}
\global\long\def\im{\mathrm{Im}}
\global\long\def\dd{\mathrm{d}}
\global\long\def\ddd{\mathcal{D}}

\global\long\def\avg#1{\left\langle #1 \right\rangle}
\global\long\def\mr#1{\mathrm{#1}}
\global\long\def\mb#1{{\mathbf #1}}
\global\long\def\mc#1{\mathcal{#1}}
\global\long\def\tr{\mathrm{Tr}}
\global\long\def\dbar#1{\stackrel{\leftrightarrow}{\mathbf{#1}}}

\global\long\def\nth{$n^{\mathrm{th}}$\,}
\global\long\def\mth{$m^{\mathrm{th}}$\,}
\global\long\def\non{\nonumber}

\newcommand{\orange}[1]{{\color{orange} {#1}}}
\newcommand{\cyan}[1]{{\color{cyan} {#1}}}
\newcommand{\teal}[1]{{\color{teal} {#1}}}
\newcommand{\blue}[1]{{\color{blue} {#1}}}
\newcommand{\yellow}[1]{{\color{yellow} {#1}}}
\newcommand{\green}[1]{{\color{green} {#1}}}
\newcommand{\red}[1]{{\color{red} {#1}}}

\global\long\def\todo#1{\cyan{{$\bigstar$ \orange{\bf\sc #1 }}$\bigstar$} }

\global\long\def\addref#1{\orange{{$\bigstar$ \cyan{\bf\sc Add reference }}$\bigstar$} }

\global\long\def\redflag#1{\Rflag{first} \red{\bf \sc #1}}

\newcommand{\ks}[1]{{\textcolor{teal}{[KS: #1]}}}

\title{Dipole-dipole Interactions Through a Lens}
\author{A. Olivera}
\email{anibal.olivera.m@gmail.com}
\affiliation{Departamento de F\'isica, Facultad de Ciencias F\'isicas y Matem\'aticas, Universidad de Concepci\'on, Concepci\'on, Chile}

\author{K. Sinha}
\email{kanu.sinha@asu.edu}
\affiliation{Department of Electrical and Computer Engineering, Princeton University, Princeton, New Jersey 08544, USA}
\affiliation{School of Electrical, Computer and Energy Engineering, Arizona State University, Tempe, AZ 85287-5706, USA}

\author{P. Solano}
\email{psolano@udec.cl}
\affiliation{Departamento de F\'isica, Facultad de Ciencias F\'isicas y Matem\'aticas, Universidad de Concepci\'on, Concepci\'on, Chile}
\begin{abstract}
We study the fluctuation-mediated interactions between two atoms in the presence of an aplanatic lens, demonstrating an enhancement in their resonant dipole-dipole interaction. We derive the field propagation of the linear optical system in terms of the electromagnetic Green's tensor for an aplanatic lens. The collective internal atomic dynamics is analyzed via a Lindblad master equation, which allows one to characterize the dispersive and dissipative interactions between atoms. We thus demonstrate that the resonant dipole-dipole coupling between the atoms can be enhanced in the focal plane of the  lens, and the lens-modified energy exchange between the atoms can  create a mutual trapping potential. Our work opens new avenues for expanding dipole-dipole interactions to macroscopic scales and the experimental platforms to study them. 
\end{abstract}

\maketitle

\section{Introduction}

Technological advances in the last decade have facilitated the probing and control of single atoms by collecting and focusing light with the help of high numerical aperture (NA) lenses. Some example of this progress are quantum gas microscopes \cite{Bakr2009, Cheuk2015,Parson2015,Haller2015,Yamamoto2016}, programmable atom arrays \cite{Endres2016,Barredo2016,Kaufman2021}, and other novel arrangements of lenses to improve atom-field interfaces \cite{Chin2017,Bianchet2021}. State-of-the-art optical elements allow for an NA as high as 0.92 \cite{Robens17}, near the theoretical limit. The rapid progress of such tools opens new possibilities to enhance and manipulate long-range atom-atom interactions.

The ability to collect light from an emitter and guide it over long distances enables a variety of collective quantum optical phenomena, which has been a subject of significant interest in recent theoretical \cite{Asenjo17, Sinha2020, Sinha2019,  Dinc19,Calajo19, Sheremet2021WaveguideQE, Trivedi21, Buonaiuto21, Poshakinskiy21, Pivovarov21, Jones20, Hughes2021} and experimental works \cite{vanLoo2013,Solano2017,Kim2018,Newman2018,Boddeti2022} in waveguide quantum electrodynamics (QED). Typically, these implementations rely on the evanescent light-matter coupling wherein the emitters are either placed nearby or embedded in a solid waveguide structure. This introduces various  dissipation and decoherence mechanisms and hinders the control and probing of both the atoms and the electromagnetic (EM) field \cite{Fermani07, Yeung96, Scheel05, Rekdal04, Skagerstam06, Sague07}.  On the contrary, imaging systems involve large distances of operation that allow one to treat the atoms as if they were in free space, while facilitating interactions with their distant counterparts.

At a fundamental level in  QED, the interactions between two atoms are mediated by the quantum fluctuations of the EM field. Such interactions depend on the range of separation between the two atoms \cite{Goldstein97, Milonni74, Sinha2020b}, boundary conditions on the EM field \cite{Dung02, Kobayashi95, Goldstein96, ElGanainy2013, Agarwal98, Hopmeier99,Haugland21} and its spectral density \cite{John95, Bay97a, Bay97b, Xie03,Kurizki90, Cortes17}, quantum correlations between the atoms \cite{Behunin10,Jones18, CollCP18,Yang2020}, external classical driving fields \cite{ Yang2020,Varada92,deLeseleuc17}, among other factors. In this work we explore the idea of using an ideal lens together with a weak external drive to amplify and engineer the  interaction between two distant atoms.  As the atoms  scatter the laser field, the lens collects and amplifies the  far-field resonant dipole-dipole interaction mediated via the drive photons. This opens the possibility of using atomic imaging technology for   engineering long-range dipole-dipole interactions and implementing collective systems without the downsides of near-field interactions.

The rest of the paper is organized as follows. In  Sec. \ref{Section: Model} we describe the system consisting of two two-level atoms placed near each focal point of an ideal, aberration free, aplanatic lens. We derive the collective atomic master equation in Sec. \ref{Section : Master Equation} and the Green's tensor for the EM field propagation in  Sec. {\ref{Section : Green's Tensor}}. This allows one to obtain the  dispersive and dissipative contributions to the effective dipole-dipole interaction in Sec. \ref{Section: Dipole-dipole interactions}. In Sec. \ref{Section : Forces}, we analyze the effects of such lens-mediated enhancement of the far-field resonant dipole-dipole interaction and the possibility of creating a mutual trap potential for atoms interacting via lenses. We finalize presenting a brief outlook and concluding remarks in Sec. \ref{Section: Summary}.

\section{Model}
\label{Section: Model}

\begin{figure}[t]
    \centering
    \includegraphics[width = 3.45 in]{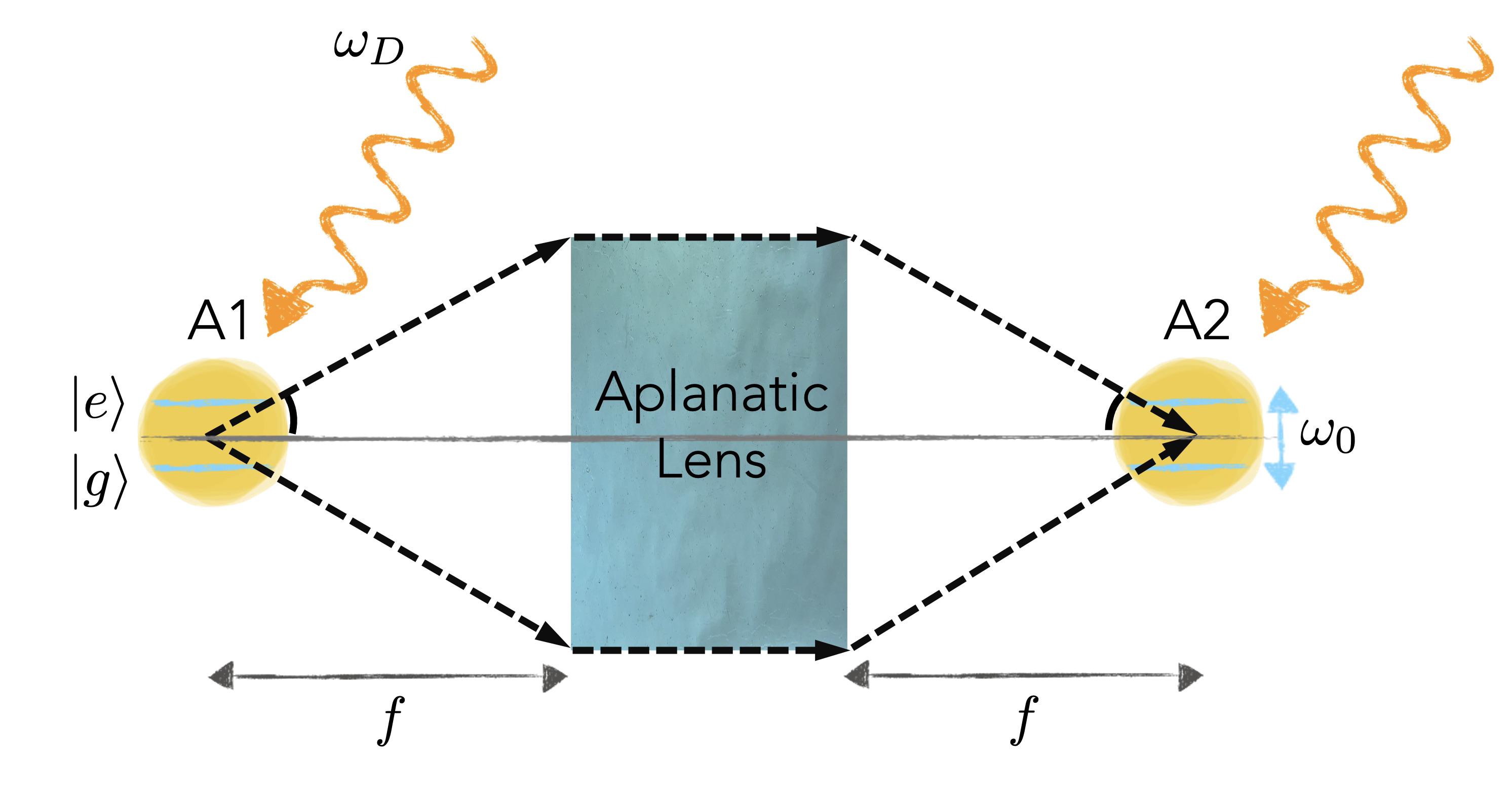}
    \caption{Schematic representation of two two-level atoms, A1 and A2,  interacting with each other via an aplanatic lens. The atoms are placed at the two focal points of the optical system at $r_1 = \cbkt{0,0,-f}$ and $r_2 = \cbkt{0,0,f}$, with $f$ as the focal length. Each atom has a resonance frequency of $\omega_0 $ and is  weakly driven by a laser of frequency $\omega_D$. }
    \label{Fig:Sch}
\end{figure}

We consider a system of two two-level atoms (A1 and A2) separated by an aplanatic lens, as shown in Fig.\,\ref{Fig:Sch}. An aplanatic lens is generally composed of two or three lenses such that spherical and coma aberrations are suppressed \cite{Pedrotti}. For the purposes of this work, we describe the lens in terms of its effects on the EM field wavefront, regardless of the details of the elements that it is comprised of. The atoms are placed near the focal points on each side of the lens. The system exhibits azimuthal symmetry, that can be broken by the polarization of the atomic dipoles  deviating from the optical axis. We further assume that the atoms are weakly driven by a classical field of frequency $\omega_D$. 

The total Hamiltonian of the system is given by $ H=H_{A}+H_{F}+H_{AF}+H_{AD}$, where $H_A$ corresponds to the Hamiltonian for the two atoms in the rotating frame with respect to the drive frequency:
\eqn{H_{A}=& \hbar \delta_D \sum_a\hat{\sigma}_{+}^{(a)} \hat{\sigma}_{-}^{(a)}}
with  $\hat{\sigma}_{+}^{(a)} \equiv \ket{e_a}\bra{g_a}$ and $\hat{\sigma}_{-}^{(a)}\equiv\ket{g_a}\bra{e_a}$ as the raising and lowering operators for the internal degrees of freedom of the atoms labeled by $a=1,2$ which corresponds to the atoms A1 and A2 respectively, and $\delta_D = \omega_{0} - \omega_D$ as the detuning between the atomic resonance  $\omega_0 $ and the drive frequency.

The atom-vacuum and atom-drive interaction Hamiltonians in the rotating frame are respectively given as:
\eqn{
    H_{A F}=&  - \sum_a\hat{\mb{p}}_a\cdot \hat{\mb{E}}\bkt{\mb{r}_a}\label{Eq:Hint},\text{and}\\
    H_{AD}=& \sum_{a}\hbar\Omega\sbkt{\sigma_{+}^{(a)}  +\sigma_{-}^{(a)}} \label{Atomic Drive Hamiltonian}.
}
The atomic dipole operator for each atom in the rotating frame is given by $\hat{\mathbf{p}}_a =\mathbf{d}_a^{\dagger}\hat{\sigma}_{+}^{(a)}e^{i \omega_D t}+\mathbf{d}_a\hat{\sigma}_{-}^{(a)}e^{-i \omega_D t}$, with $\mathbf{d}_a$ the dipole matrix element associated with the $\ket{g_a}\leftrightarrow\ket{e_a}$ transition.  $\hat{\mb{E}}\bkt{\mb{r}_a}$ represents the electric field at position $\mb{r}_a$ of atom $a$. The  Rabi frequency of the drive is given by $\Omega$. We note that in the presence of a weak classical drive, the atomic dipoles exhibit Rayleigh scattering at the drive frequency.

The field Hamiltonian $H_F $ and the quantized EM field in the presence of media are described in the macroscopic QED formalism \cite{Gruner96, BuhmannRev, Dung02, Buhmann1, Buhmann2}, as discussed in Appendix\,\ref{App:MQED}.

\section{Atomic Master equation}

\label{Section : Master Equation}
We can now describe the dynamics of the atomic internal degrees of freedom in terms of a second-order Lindblad master equation by tracing out the EM field in the Born-Markov approximations (see Appendix\,\ref{App:ME} for details) \cite{BPbook, CollCP18}:
\eqn{\label{Eq:ME}
\der{\rho_A}{t} = -\frac{i}{\hbar}\sbkt{H_A', \rho_A} + \mc{L}_A \sbkt{\rho_A},
}
where $\rho_A $ corresponds to the collective density matrix of the two atoms. The effective Hamiltonian   $H_A' $  and the Liouvillian  $ \mc{L}_A$ describe the dispersive and the  dissipative dynamics of the collective atomic system in the presence of the aplanatic lens: 
\eqn{
    H_A' = &\sum_{i,j = 1,2} J^{(+)}_{ij} \hat\sigma_+^{(i)} \hat\sigma_-^{(j)}+J^{(-)}_{ij}\hat \sigma_-^{(i)}\hat \sigma_+^{(j)}\label{Atomic Hamiltonian Traced},\\
    \mathcal{L}_{A}\sbkt{\rho_A} =& -\frac{1}{2}\sum_{i, j = 1,2} \Gamma_{ij}\cbkt{ \hat{\sigma}_{+}^{(i)} \hat{\sigma}_{-}^{(j)}, \rho_A}  \non\\
    &+
    \sum_l\int d^{3} \mathbf{k}\hat{ \mc{O}}^{(i)}_{\mb{k}l}  \rho_A \bkt{\hat{\mc{O}}^{(j)}_{\mb{k}l}}^\dagger.
}
The coherent couplings  between the two atoms and the individual energy modifications to the excited and ground states of the atoms are given by  $J_{ij}^{(+)}=- J_{ij}^\mr{OR} -J_{ij}^\mr{R}$ and $J_{ij}^{(-)} = J_{ij}^\mr{OR}$. The off-resonant and the resonant contributions $J_{ij}^\mr{OR}$ and $J_{ij}^\mr{R}$ correspond to the contributions from virtual and real photons, respectively, and are given explicitly as follows:
\eqn{
\label{Eq:JOR}
J_{ij}^\mr{OR}& \equiv\frac{\mu_0 \omega_D}{\pi} \int_0^\infty \dd\xi\, \frac{\xi^2}{\xi^2+\omega_D^2}\bkt{{\bf d}^\dagger\cdot {\dbar{G}\bkt{{\bf r}_i, {\bf r}_j,i\xi}}\cdot{\bf d}}\\
 \label{Eq:JR}
J_{ij}^\mr{R} &\equiv \mu_0 \omega_D^2 \re\sbkt{{\bf d}^\dagger\cdot {\dbar{G}\bkt{{\bf r}_i, {\bf r}_j, \omega_D}}\cdot{\bf d}}.
} We note that while the off-resonant part depends on the broadband frequency response of the environment, the resonant part only depends on the response of the EM environment at the drive frequency. 

The dissipative interaction between the atoms is given by:
\eqn{
\label{Eq:gammaij}
\Gamma_{ij}\equiv&\frac{2\mu_0\omega_D^2 }{\hbar} \, {{\bf d}^\dagger\cdot \im\sbkt{\dbar{G}\bkt{{\bf r}_i, {\bf r}_j, \omega_D}}\cdot{\bf d}},
}
which is related to the resonant dispersive interaction $\bkt{J_{ij}^\mr{R}}$ via the Kramers-Kronig relation \cite{Buhmann2}.

The jump operator $\hat{\mc{O}}_{\mb{k}, l}^{(i)}$ for atom $i$ \cite{Chang14}:
\eqn{\label{Eq:jump}
&\mc{\hat{O}}_{\mb{k},l}^{(i)}=\non\\
&\sqrt{ \frac{2\epsilon_0 \mu_0 ^2 \omega_D^4}{\hbar }} \int\dd^3 r \frac{e^{i \mb{k}\cdot\mb{r}}}{\bkt{2\pi}^{3/2} }\sqrt{\epsilon\bkt{\mb{r},\omega_D}}\mb{d}_i{G}_{il}\bkt{\mb{r}_i,\mb{r},\omega_D} \hat{\sigma}^{(i)}_-,
}
corresponds to the process of  recoil of a photon of frequency $\omega_D$, momentum $k$ and polarization  $l$; similarly for atom $j$. It can be seen that $ \sum_{l}\int\dd^3 \mb{k}\bkt{\hat{\mc{O}}^{(i)}}^\dagger_{\mb{k},l}\hat{\mc{O}}^{(j)}_{\mb{k},l} = \hbar \Gamma_{ij} \hat{\sigma}^{(i)}_+\hat{\sigma}^{(j)}_-$.

 When analyzing the far-field contributions to the dipole-dipole interactions amplified by the lens we can neglect the off-resonant contributions from virtual photons at second-order $(J_{ij}^\mr{OR})$ that scale as $\sim 1/r^{3} $, and those at fourth-order that scale as $\sim 1/r^{6}$. The predominant contribution is the resonant dipole-dipole interaction given by Eq.\,\eqref{Eq:JR}, which scales as $\sim 1/r$, and depends only on the response of the environment at the drive frequency $\omega_D$.

\section{Green's Tensor for the  Lens}
\label{Section : Green's Tensor}

We describe the electric field at a given point $\mb{r}$ emitted by a dipole at the position $\mb{r}_0$ in terms of the classical EM Green's tensor \cite{Buhmann1,Buhmann2}, defined by the inhomogeneous Helmholtz equation as follows:
\eqn{
\sbkt{\frac{1}{\mu_0}\mb{\nabla}\times\mb{\nabla}\times - \frac{\omega^2}{c^2} \epsilon\bkt{ \mb{r,\omega}}} \dbar{G} \bkt{\mb{r},\mb{r}_0,\omega} = \delta \bkt{\mb{r} - \mb{r}_0},
}
together with the condition that $\dbar{G}\bkt{\mb{r},\mb{r}_0, \omega}\rightarrow0$ as $ \abs{\mb{r} - \mb{r}_0}\rightarrow\infty$. One can separate the contributions to the field in terms of the distance between the the two points as:
\eqn{\stackrel{\leftrightarrow}{\mathbf{G}}\bkt{\mb{r},\mb{r}_0, \omega} =& \stackrel{\leftrightarrow}{\mathbf{G}}_{\mathrm{NF}}\bkt{\mb{r},\mb{r}_0, \omega}+\stackrel{\leftrightarrow}{\mathbf{G}}_{\mathrm{IF}}\bkt{\mb{r},\mb{r}_0, \omega}\non\\
&+\stackrel{\leftrightarrow}{\mathbf{G}}_{\mathrm{FF}}\bkt{\mb{r},\mb{r}_0, \omega},\label{FullFreeGreenTensor}}
where $\dbar{G}_\mr{NF}\bkt{\mb{r},\mb{r}_0, \omega}$, $\dbar{G}_\mr{IF}\bkt{\mb{r},\mb{r}_0, \omega}$ and $\dbar{G}_\mr{FF}\bkt{\mb{r},\mb{r}_0, \omega}$ correspond to the near-field, intermediate-field and far-field contributions, respectively. Since the distance between the emitter and the lens is much greater than the wavelength of the emitted field, we will only be interested in the far-field contribution \cite{Buhmann1,Buhmann2,Novotnybook}.

\begin{figure}[t]
    \centering
    \includegraphics[width=3.5 in]{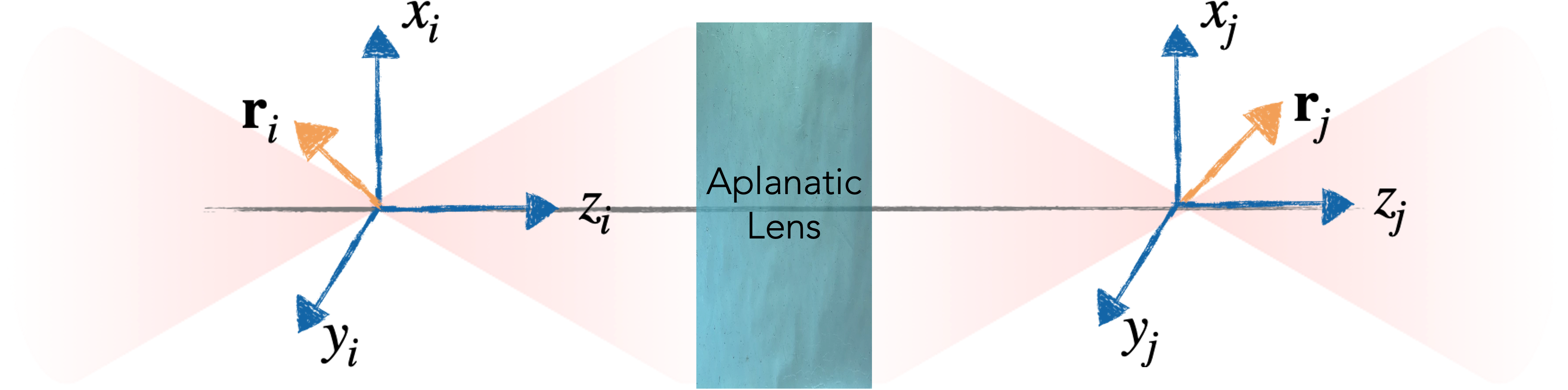}
    \caption{ Coordinate systems for the linear optics configuration of Fig. \ref{Fig:Sch}. Each end has an origin at its respective origin from which the positions $\mb{r}_i$ and $\mb{r}_j$ arise. Within the focal zone the intensities only depend on the relative values $\stackrel{\leftrightarrow}{\mathbf{G}}_{\mathrm{PSF}}\bkt{\mb{r}_{i},\mb{r}_{j}, \omega_D}=\stackrel{\leftrightarrow}{\mathbf{G}}_{\mathrm{PSF}}\bkt{x_{ij},y_{ij},z_{ij},\omega_D}$.}
    \label{fig:CoordinateSystem}
\end{figure}

Let $\mb{r}_1$ and $\mb{r}_2$ be the positions of the atoms A1 and A2, as depicted in Fig. \ref{fig:CoordinateSystem}. The resonant electric field produced by an arbitrarily oriented electric dipole located at $\mb{r}_1$, with dipole moment $\mathbf{p}$, after propagation to a point $\mathbf{r}$ at the vacuum-lens interface is:

\begin{equation}\label{Far Field Lens}
    \mathbf{E}_{\mathrm{FF}}\bkt{\mb{r},\mb{r}_{1}} = - \frac{\omega^{2}_D}{\varepsilon_{0} c^{2}}\stackrel{\leftrightarrow}{\mathbf{G}}_{\mathrm{FF}}\bkt{\mb{r},\mb{r}_{1}, \omega_D}\cdot \mathbf{p}.
\end{equation}
The field is collimated after transmission through the first interface of the aplanatic lens, travels as a collimated beam until it reaches the second interface on the opposite side after which it converges at the focal point on the other side.  To obtain the field in the region near the second focus, we use the \textit{angular spectrum representation} in cylindrical coordinates $\cbkt{\rho,\varphi,z}$ \cite{Novotnybook}:
\begin{align}
    \mathbf{E}_{\mathrm{fo}}(\rho, \varphi, z)&=-\frac{\mathrm{i} k_Df \mathrm{e}^{-\mathrm{i} k_D f}}{2 \pi}  \int_{0}^{\theta_{\max }} \mathrm{d} \theta\,\sin \theta  \nonumber \\ 
     &\int_{0}^{2 \pi}  \mathrm{d} \phi  \ \mathbf{E}_{\mathrm{FF}}(\theta, \phi)\mathrm{e}^{\mathrm{i} k_D z \cos \theta} \mathrm{e}^{\mathrm{i} k_D \rho \sin \theta \cos (\phi-\varphi)} \label{AngularSpectrumRepresentation}
\end{align}
where $k_D=\omega_D/c$, and the coordinate $\theta_{\max}$ represent the maximum angle of a cone of light that can enter the lens, determined by the NA of the lens (NA$=\sin \theta_{\max}$, in vacuum). Substituting  Eq.\,\eqref{Far Field Lens} in Eq.\,\eqref{AngularSpectrumRepresentation} gives a compact form for the electric field in the focal zone $\mb{r}_{2}=\cbkt{\rho,\varphi,z}$ (see Appendix \ref{App: Green Tensor of a Lens}): 
\begin{equation}\label{FocalField}
    \mathbf{E}_{\mathrm{fo}}\bkt{\mb{r}_{2},\mb{r}_{1}} = \frac{\omega^{2}_D}{\varepsilon_{0} c^{2}} \stackrel{\leftrightarrow}{\mathbf{G}}_{\mathrm{PSF}}\bkt{\mb{r}_{2},\mb{r}_{1}, \omega_D} \cdot \mathbf{p},
\end{equation}
where $\stackrel{\leftrightarrow}{\mathbf{G}}_{\mathrm{PSF}}\bkt{\mb{r}_{1},\mb{r}_{2}, \omega_D}$ is the \textit{dyadic point-spread function} of the optical system with equal focal lengths $f$ on each side of the lens, which describes the field propagation between both focal regions. In general, Eq.\,\eqref{FocalField} connects atoms at opposite ends with the point spread function given by:
\eqn{
\label{GreenPSF}
    &\stackrel{\leftrightarrow}{\mathbf{G}}_{\mathrm{PSF}}\bkt{\mb{r}_{i},\mb{r}_{j}, \omega_D} = \frac{\omega_D}{8 \pi c}  \dbar{\mb{g}}\bkt{\mb{r}_i, \mb{r}_j, \omega_D},
}
where we have defined
\eqn{&\dbar{\mb{g}}\bkt{\mb{r}_i, \mb{r}_j, \omega_D} =\mathrm{i}\non\\
&\bkt{\begin{array}{ccc}
    I_{1}+I_{2} \cos \left(2 \varphi_{i j}\right) & I_{2} \sin \left(2 \varphi_{i j}\right) & -2 \mathrm{i} I_{3} \cos \left(\varphi_{i j}\right) \\
    I_{2} \sin \left(2 \varphi_{i j}\right) & I_{1}-I_{2} \cos \left(2 \varphi_{i j}\right) & -2 \mathrm{i} I_{3} \sin \left(\varphi_{i j}\right) \\
    -2 \mathrm{i} I_{3} \cos \left(\varphi_{i j}\right) & -2 \mathrm{i} I_{3} \sin \left(\varphi_{i j}\right) & 2 I_{4}
    \end{array}}
    }
The integrals $I_n$ are 
\begin{align}
    I_1 =& \int_{0}^{\theta_{\text{max}}}    \dd\theta \sin\theta \sbkt{ 1+\cos^2 \theta }  \mathrm{e}^{\mathrm{i}k_{D}\cos\theta|z_{ij}|} J_0\bkt{k _D\rho_{ij}\sin \theta} \label{Integral_I1}
    \\
    I_2 =& \int_{0}^{\theta_{\text{max}}}  \dd\theta \sin\theta \sbkt{ 1-\cos^2 \theta }  \mathrm{e}^{\mathrm{i}k_{D}\cos\theta|z_{ij}|} J_2\bkt{k _D\rho_{ij}\sin \theta} \label{Integral_I2}
    \\
    I_3 =& \int_{0}^{\theta_{\text{max}}}  \dd\theta \sin^2\theta \cos\theta \mathrm{e}^{\mathrm{i}k_{D}\cos\theta|z_{ij}|} J_1\bkt{k_D\rho_{ij}\sin \theta} \label{Integral_I3}
    \\
    I_4 =& \int_{0}^{\theta_{\text{max}}}  \dd\theta \sin^3\theta  \mathrm{e}^{\mathrm{i}k_{D}\cos\theta|z_{ij}|}  J_0\bkt{k_D\rho_{ij} \sin \theta} , \label{Integral_I4}    
\end{align}
where $J_{n}$ is the $n^\text{th}$ order Bessel functions of the first kind, and the relative coordinates between the atoms at each end are given by $\rho_{ij}=\sqrt{x_{ij}^2+y_{ij}^2}$, $\tan\varphi_{ij}=y_{ij}/x_{ij}$, $x_{ij}=x_i-x_j$, $y_{ij}=y_i-y_j$, and $z_{ij}=z_i-z_j$. The symmetry under the permutation $i \leftrightarrow j$ (or equivalently for one atom on each side, change A1 for A2) allows one to deduce that for $\mb{r}_i$ and $\mb{r}_j$ in the focal zone, only the relative distance between the two points will produce detectable changes. The change from absolute to relative perspective is evidenced in the fulfillment of the Onsager reciprocity \cite{Buhmann1} for symmetric tensors
\eqn{
    \stackrel{\leftrightarrow}{\mathbf{G}}_{\mathrm{PSF}}\bkt{\mb{r}_{i},\mb{r}_{j}, \omega_D} = \stackrel{\leftrightarrow}{\mathbf{G}}_{\mathrm{PSF}}\bkt{\mb{r}_{j},\mb{r}_{i}, \omega_D}.
}

Since $f \gg \lambda_D$, it is only necessary to consider the far-field term $\stackrel{\leftrightarrow}{\mathbf{G}}_{\mathrm{FF}}$ of the full free-space Green's tensor Eq.\,\eqref{FullFreeGreenTensor} for the derivation of Eq.\,\eqref{GreenPSF}.  Thus, $\stackrel{\leftrightarrow}{\mathbf{G}}_{\mathrm{PSF}}$ accounts for the radiation collected by the lens and redirected to the other atom.

\section{Dipole-dipole interaction Lensing}
\label{Section: Dipole-dipole interactions}

Having  obtained the Green's tensor for the lens $\stackrel{\leftrightarrow}{\mathbf{G}}_{\mathrm{PSF}}\bkt{\mb{r}_{i},\mb{r}_{j}, \omega_D}$ in Eq. (\ref{GreenPSF}), the dispersive and dissipative resonant dipole-dipole coupling coefficients  between the two atoms ($J_{12}^\mr{R}$  and $ \Gamma_{12}$)  described by Eq. (\ref{Eq:JR}) and (\ref{Eq:gammaij}) respectively can be simplified as:
\eqn{ J_{12}^\mr{R}/ \bkt{\hbar \Gamma}= &
\frac{3}{8}\re\sbkt{\mb{u}_1\cdot \dbar{\mb{g}}\cdot\mb{u}_2 } \label{Eq:J}\\
    \Gamma_{12}/\Gamma =& \frac{3}{4}\im\sbkt{\mb{u}_1\cdot \dbar{\mb{g}}\cdot\mb{u}_2 } \label{Eq:Gamma},
}
where we have defined $ \Gamma \equiv \frac{\abs{\mb{d}}^2\omega_D^3}{3\pi\hbar  \epsilon_0 c^3}$ as the emission rate for a dipole radiating at the drive frequency $\omega_D$, the unit vectors $\mb{u}_j$ correspond to the orientation of the atomic dipole $j$.

\begin{figure}[]
\centering
\includegraphics[width = 3.4 in]{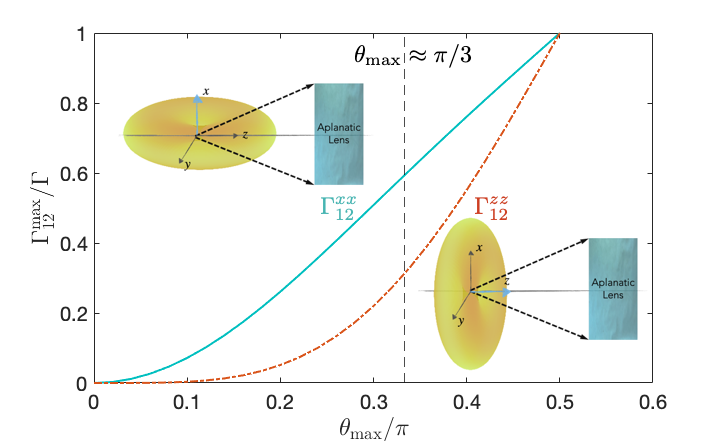}
\caption{Maximum dissipative dipole-dipole coupling ($\Gamma_{12}^\mr{max}$) as a  function of the angular aperture $\theta_\mr{max}$. The solid (dash-dotted) curve represents $\Gamma_{12}^\mr{max}$ for two $x$($z$)-oriented dipoles, as illustrated in the schematic inset figures. The dashed vertical line indicates an angular aperture of $\theta_\mr{max}\approx\pi/3$, where   $\Gamma_{12}^\mr{max}\approx0.6\Gamma$.}
\label{Fig:PatternAndCapturedEmission}
\end{figure}

The contribution of the lens to the enhancement of the dipole-dipole interactions can be  characterized by the maximum dissipative dipole-dipole interaction ($\Gamma_{12}^\mr{max}$) between two dipoles placed at the foci. Fig. \ref{Fig:PatternAndCapturedEmission} shows $\Gamma_{12}^\mr{max}$ as a function of the NA for two orthogonal orientations of the atomic dipole. We see that for a feasibly high NA of $\theta_\mr{max} \approx \pi/3$ and appropriate atomic dipole alignment, the dipole-dipole interaction rate can reach nearly 60\% of the atomic decay rate.

\begin{figure*}
    \centering
    \includegraphics[width = 5 in]{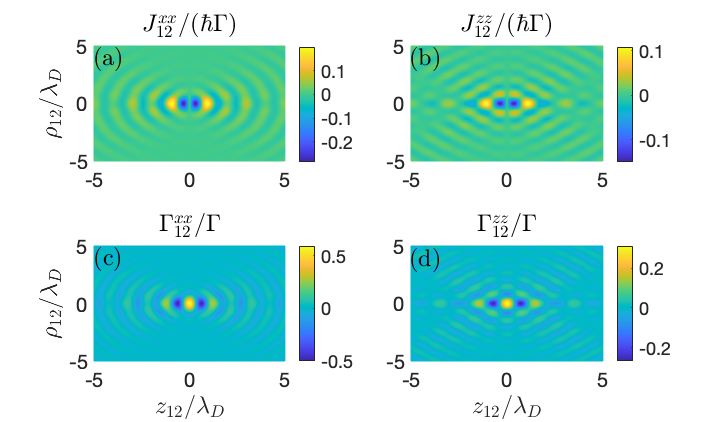}
    \caption{Spatial distribution of  the (a,b)  dispersive  and (c,d)  dissipative  interaction in the $xz-$plane. The presence of an atom A1 emitting radiation produces a lensed field in the focal zone at the opposite end. A second atom A2 at this end will be subject to dispersive and dissipative interactions depending on the relative positions, correlations  and dipole orientations of the two atoms. We have chosen an angular aperture of  $\theta_{\text{max}}=\pi/3$ in the figures above.}
    \label{Fig:GammaJ}
\end{figure*}

Fig.\,\ref{Fig:GammaJ}  shows a the spatial dependence of the resonant dipole-dipole coupling $J_{12}^{\text{R}}$ ((a) and (b)) and $\Gamma_{12}$ ((c) and (d)) for dipoles with orthogonal ($x$) and parallel ($z$) orientations with respect to the optical axis, evidencing a lensing effect near the focal zone of one of the dipoles. The fringes with periodicity $\sim \lambda_D$ correspond two constructive and destructive interference effects in the collective dipole-dipole interactions, leading to super- and sub-radiant dispersive and dissipative interactions.

\section{Dipole-dipole potential through the lens}
\label{Section : Forces}

\begin{figure}[b]
    \centering
    \includegraphics[width = 2.5 in ]{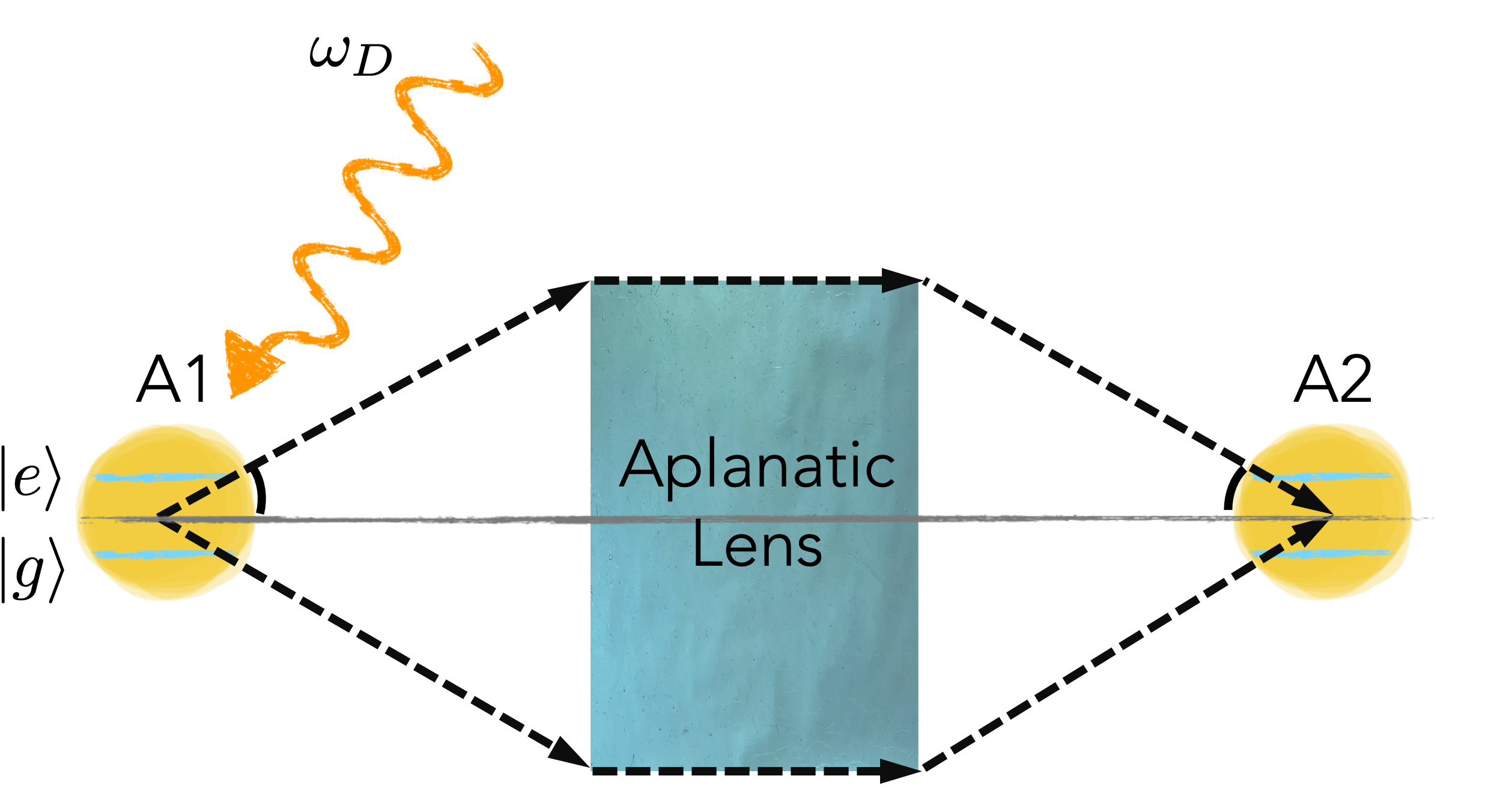}
    \caption{Schematic representation of the system of two atoms A1 and A2 interacting via the lens with A1 being driven. The dipole-dipole interaction between the atoms creates a trap-like potential for atom A2.}
    \label{Fig:Sch CP}
\end{figure}

The lens-mediated energy exchange between the atoms can be significant enough to create a mutual trapping potential. Let us consider the scenario depicted in Fig. \ref{Fig:Sch CP} where a single (trapped) atom A1 is externally driven and coupled to a second atom A2 through the lens. Following Eq.\,\eqref{Atomic Hamiltonian Traced}, the steady-state potential energy created by the exchange of real photons between atoms A1 and A2 is given by
\begin{equation}\label{Eq:Expected Value Ham}
    \left\langle H_{A}^{\prime}\right\rangle_{\mr{ss}} = -J^{\text{R}}_{12}\xi,
\end{equation}
where $\xi=\left\langle \sigma_{+}^{(1)}\sigma_{-}^{(2)}\right\rangle_{\mr{ss}}+\left\langle \sigma_{+}^{(2)}\sigma_{-}^{(1)}\right\rangle_{\mr{ss}} $ represents the atomic cross-correlations.

We can compute the steady state solutions of the internal atomic state via the equations of motion for the operators $\hat{\sigma}_{-}^{(j)}$ in the Heisenberg picture, considering Eq.\,\eqref{GreenPSF} and Eq.\,\eqref{Eq:ME}. In the low saturation approximation, this yields:
\begin{align}
    \dot{\hat{\sigma}}_{-}^{(1)}=&\left(\mathrm{i}\delta_{D}-\frac{\Gamma}{2}\right)\hat{\sigma}_{-}^{(1)}+\mathrm{i}\Omega \nonumber +\mathrm{i}G_{12}\hat{\sigma}_{-}^{(2)}\\
    \dot{\hat{\sigma}}_{-}^{(2)}=&-\frac{\Gamma}{2}\hat{\sigma}_{-}^{(2)}+\mathrm{i}G_{12}\hat{\sigma}_{-}^{(1)},
\end{align}
where $\delta_D$ and $\Omega$ are the detuning and Rabi frequency of the external drive and we define $G_{12}\equiv\frac{\mu_{0}\omega_{D}^{2}}{\hbar}\mathbf{d}^{\dagger}\cdot \stackrel{\leftrightarrow}{\mathbf{G}}_{\text {PSF }}\left(\mathbf{r}_{1}, \mathbf{r}_{2}\right)\cdot\mathbf{d}\equiv J_{12}/\hbar+\mr{i}\Gamma_{12}/2$ as the effective complex dipole-dipole coupling. Solving for the steady-state expectation value of the atomic operators and their correlations we obtain: 
\begin{align}
    \left\langle \sigma_{-}^{(1)}\right\rangle_{\mr{ss}} =&\frac{-\mr{i}\Omega}{\mathrm{i}\delta_D-\Gamma/2-2G_{12}^{2}/\Gamma}\label{Eq:steady-state 1} \\
    \left\langle \sigma_{-}^{(2)}\right\rangle_{\mr{ss}} =&2\mr{i}\frac{G_{12}}{\Gamma}\left\langle \sigma_{-}^{(1)}\right\rangle_{\mr{ss}} \label{Eq:steady-state 2}\\
    \left\langle \sigma_{+}^{(1)}\sigma_{-}^{(1)}\right\rangle_{\mr{ss}} =&-\frac{2}{\Gamma}\operatorname{Im}\left[G_{12}\left\langle \sigma_{+}^{(1)}\sigma_{-}^{(2)}\right\rangle_{\mr{ss}} \right] \nonumber\\ &+\frac{2\Omega}{\Gamma}\operatorname{Im}\left[\left\langle \sigma_{-}^{(1)}\right\rangle_{\mr{ss}} \right].
    \label{Eq:rho_ee_1}\\
    \left\langle \sigma_{+}^{(2)}\sigma_{-}^{(2)}\right\rangle_{\mr{ss}} =&\frac{2}{\Gamma}\operatorname{Im}\left[G_{12}^{*}\left\langle \sigma_{+}^{(1)}\sigma_{-}^{(2)}\right\rangle_{\mr{ss}} \right] \label{Eq:rho_ee_2}
\end{align}

The steady-state equations lead to the expected result of driving just A1 when the dipole-dipole coupling vanishes $(G_{12}\rightarrow0)$. Otherwise, the  dispersive shift and linewidth of A1 are modified by the lens-mediated interaction between A1 and A2, as Eq.\,\eqref{Eq:steady-state 1} shows. Furthermore, the probabilities of finding each atom in an excited state, given by Eqs.\,\eqref{Eq:rho_ee_1} and \eqref{Eq:rho_ee_2}, depend on the dipole-dipole coupling and the strength of the atomic cross-correlations
\begin{equation}\label{Eq:Correlations}
    \left\langle \sigma_{+}^{(1)}\sigma_{-}^{(2)}\right\rangle_{\mr{ss}} =\frac{\alpha\beta+2\left|G_{12}\right|^{2}\beta^{*}}{\left|\alpha\right|^{2}-4\left|G_{12}\right|^{4}},
\end{equation}
where
\begin{align}
    \alpha=&2\re \left[G^{2}_{12}\right]+\Gamma(\Gamma-\mr{i}\delta_{D}), \nonumber\\
    \beta=&\Omega G\left\{3\left\langle \sigma_{-}^{(1)}\right\rangle_{\mr{ss}} -\left\langle \sigma_{+}^{(1)}\right\rangle_{\mr{ss}} \right\}.
\end{align}
Assuming the low saturation limit, one gets that $\left\langle \sigma_{+}^{(1)}\sigma_{-}^{(2)}\right\rangle_{\mr{ss}}\approx \left\langle \sigma_{+}^{(1)}\right\rangle_{\mr{ss}}\left\langle\sigma_{-}^{(2)}\right\rangle_{\mr{ss}}$  \cite{Eldredge,Chang2013}, which can be  numerically verified. This corresponds to a semiclassical limit in which the atoms radiate as classical antennas.

All atomic populations and correlations are ultimately a function of the probability of A1 being in the excited state. The saturation parameter for A1 in the absence of the lens-mediated coupling ($G_{12}=0$) is given by 
\begin{equation}\label{Eq:Saturatoion parameter}
    s   =\abs{\avg{\sigma_-^{(1)}}}_{G_{12}=0}=  \sqrt{\frac{\Omega^2}{\delta_D^2+\Gamma^2/4}},
\end{equation}
and $s^2$ is the probability of finding A1 in the excited state. To compare the response of the system at different driving frequencies on an equal footing we fix the saturation parameter $s$, meaning that we have to adjust the intensity of the drive as $\delta_D$ changes, setting  
$\Omega = s\sqrt{\delta_D^2 + \Gamma^2/4}$. 
 
We now analyze the steady state solutions for a suitable trapping configuration, focusing on two atoms oriented parallel to the x-axis of the coordinate system established in Fig. \ref{fig:CoordinateSystem}. In order to have an attractive potential for the atom we must look for zones of maximum $J_{12}$ in Fig. \ref{Fig:GammaJ} (a), where we can see that in the vicinity of $z^{\text{min}}_{12}\approx 0.92 \lambda_D$ there is a suitable trapping potential. In such position, $\Gamma^{\text{min}}_{12}\approx-0.15\Gamma$ and $J^{\text{min}}_{12}\approx0.4\hbar\Gamma$, which we use to define $G_{12}^{\rm{min}}$. We consider this particular conditions to analyze the trapping capabilities of the system.

\begin{figure}[t]
    \centering
    \includegraphics[width=0.9\linewidth]{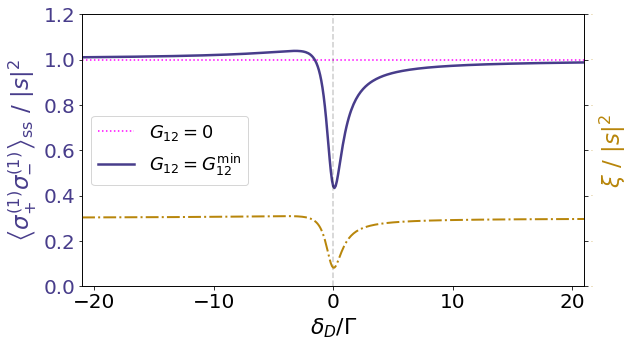}
    \caption{Left axis: Probability of finding A1 in the excited state as a function of $\delta_D$, with (solid purple) and without (dotted pink) dipole-dipole interaction with atom A2. Right axis:  atomic cross-correlations $\xi$ (dashed gold) as a function of the detuning. All the quantities are normalized by the $s^2$ to compare them with the probability of exciting A1 in the absence of dipole-dipole interactions.}
    \label{Fig:rho_11}
\end{figure}

Fig. \ref{Fig:rho_11} illustrates the effect of lens-mediated dipole-dipole interaction on the atom A1 and the  cross-correlations, $\xi$. In the absence of a lens and constant saturation parameter, the probability of observing A1 to be excited is constant at all driving frequencies. When dipole-dipole interactions are present, one can see a strong excitation transfer to A2 near resonance, evidenced by a dip with an asymmetry around $\delta_D=0$ due to the contribution of $\im{G^2_{12}=J_{12}\Gamma_{12}/\hbar}$ in the denominator of Eq.\,\eqref{Eq:steady-state 1}. The  atomic cross-correlation, also shown in Fig. \ref{Fig:rho_11}, is affected in a similar way. More importantly, it is always positive, making suitable to induce a dipole-dipole trapping potential as suggested by Eq. (\ref{Eq:Expected Value Ham}).

\begin{figure*}[t]
    \centering
    \subfloat[]{\includegraphics[width= 2 in]{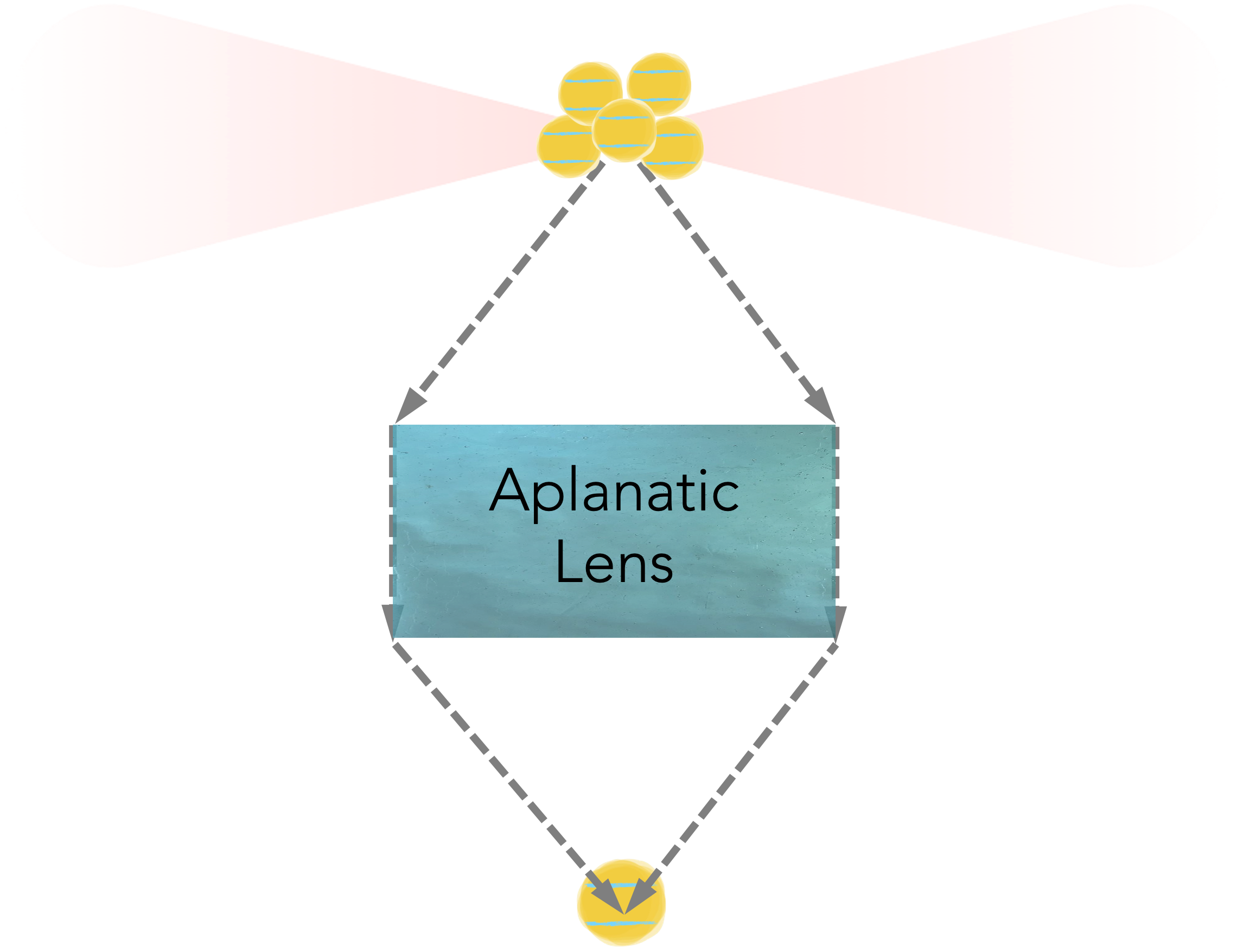}}
    \subfloat[]{
    \includegraphics[width= 4 in]{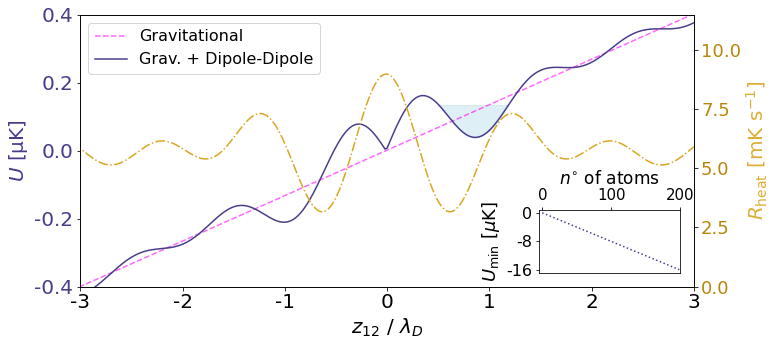}}
    \caption{(a)  Schematic representation of the trap formed by the lensed dipole-dipole  forces: $N$ atoms on the top are trapped by a tweezer at one of the focal points of the aplanatic lens system. The dipole-dipole forces between an atom placed around the other focal point of the lens and the collection of atoms on the top can be sufficiently strong to counteract gravity. (b) Left axis: comparison of potential energy from gravity (dashed pink) and dipole-dipole interaction through a lens (solid purple) evaluated in the steady state from Eq.\,\eqref{Eq:Expected Value Ham}.  The height of the blue shaded area is $E_r$, comparing the trap depth to the recoil energy.
   Right axis: Scattering rate from the emission of photons (dashed gold) obtained via Eq.\,\eqref{Heating Rate}. The inset shows the minimum potential produced by the lensing field of $0<N<200$ atoms.
   }
    \label{fig:Potential_plot}
\end{figure*}

We can estimate the average lifetime $t_{\mr{trap}}$ of the alleged atomic trap for the non-driven atom A2 by comparing the depth of the potential well (see Fig. \ref{fig:Potential_plot}) with the heating rate of A2 due to spontaneous emission. Assuming that the atom gains recoil energy after every cycle of spontaneous emission, the heating rate is given by
\begin{equation}\label{Heating Rate}
    R_{\mr{heat , pw}}^{(2)} \approx \mr{E}_{\text{r}} \Gamma_{\text{tot}}\left\langle \sigma_{+}^{(2)}\sigma_{-}^{(2)}\right\rangle_{\mr{ss}},
\end{equation}
where $\mr{E}_{\text{r}} =\hbar^2 k^2_D/2m$ is the recoil energy and $\Gamma_{\text{tot}}\approx\Gamma+\Gamma_{12}\frac{\left\langle \sigma_{+}^{(1)}\sigma_{-}^{(2)}\right\rangle_{\mr{ss}}}{\left\langle \sigma_{+}^{(2)}\sigma_{-}^{(2)}\right\rangle_{\mr{ss}}}$ is the total atomic decay rate. For the chosen $\Gamma_{12}^{(\rm{min})}$, and considering Eqs. \eqref{Eq:rho_ee_2} and \eqref{Eq:Correlations}, $\Gamma_{\text{tot}}\approx 0.93 \Gamma$, meaning slightly subradiant. The estimated trapping lifetime is
\begin{equation}\label{Eq:DeltaU_R}
    t_{\mr{trap}}=\frac{\Delta U_{\mr{pw}}}{R_{\text {heat}, \mr{pw}}^{(2)}} =\frac{\Delta J_{12}}{E_{r}}\frac{\Gamma}{\Gamma_{\text{tot}}}\frac{\im{G_{12}}}{\left|G_{12}\right|^2},
\end{equation}
where $\Delta J_{12}=J^{\mr{top}}_{12}-J^{\mr{min}}_{12}+E_{0}$, $J^{\mr{top}}_{12}$ is the value of the energy shift at the top of the potential well, and $E_{0}$ is the initial energy of the atom in the trap. We can provide a phenomenological upper bound for the expression considering the potential depth $\Delta J_{12}$ being of the order of $2\hbar\re{G_{12}}$. The term $\re{G_{12}}\im{G_{12}}/|G_{12}|^2\leq1/2$, leading to $t_{\mr{trap}}\lesssim \frac{1}{\omega _{r}}\frac{\gamma}{\Gamma_{\text{tot}}}$. This directly relates the optimum trapping lifetime to the inverse of the recoil frequency $\omega_r=E_r/\hbar$, giving an estimate of the time scale.

We now study the behavior of the trap in a realistic scenario with alkaline atoms. Let us consider ${}^{133}$Cesium atoms in and their $6^{2} \mathrm{~S}_{1 / 2} \longrightarrow 6^{2} \mathrm{P}_{3 / 2}$ transition as a two-level system, with dipole moment $\mathbf{d}=2.69 \times 10^{-29} \mathrm{C\cdot m}$, decay rate $\Gamma = 2\pi \cdot 5.23 \mathrm{MHz}$, $\lambda_{0}=852 \mathrm{nm}$ and $m=1.66 \times 10^{-27}\mathrm{Kg}$  \cite{CesiumData}. We will consider the limit $|\delta_D|\gg \Gamma$, where the system effectively behaves as a far-detuned optical dipole trap driving A1, such that A1 is trapped in a far-detuned optical dipole trap via the external drive. As a consequence, A2 is then confined only due to the interaction with A1 mediated by an aplanatic lens with an angular aperture $\theta_{\text{max}}=\pi/3$.

Figure \ref{fig:Potential_plot} shows the trapping potential $\left\langle H_{A}^{\prime}\right\rangle_{\mr{ss}}+U_g$, where $U_g$ the gravitational potential for the atom with respect to $z_{12}=0$, and heating rate as a function of the position along the optical axis. The shaded area shows the size of $E_{\mr{r}}$ compared to the potential well. We focus on the local minimum of the potential, $z^{\mr{min}}_{12}$, suitable for trapping. We can estimate the lifetime of the trap from Eq. (\ref{Eq:DeltaU_R}) assuming that the atom starts with an initial energy $E_r$ from the bottom of the trap and considering  $ J^{\rm{max}}_{12}=0.5 \hbar\Gamma$ for the present configuration. Eq. \eqref{Eq:DeltaU_R} gives us a trapping time of about $t_{\mr{trap}} \approx 1170 \gamma^{-1}_0 (\approx 1/2\omega_r)$.

As we see from Fig.\ref{fig:Potential_plot}, the size of the potential well created by the dipole-dipole interactions with a single atom is of the order of the recoil energy, possibly making trapping impractical. However, we can increase the trapping potential by increasing the number of atoms being driven. The general interaction Hamiltonian $H^{\prime}_{A}$ with $N_i$ driven atoms on the top of the optical system shown in Fig. \ref{Fig:Sch CP} is
\begin{equation}
    H_{A}^{\prime}= -\sum_{i}^{N_{i}} J^{\text{R}}_{i, A 2}\left(\hat{\sigma}_{+}^{(i)} \hat{\sigma}_{-}^{(2)}+\hat{\sigma}_{+}^{(2)} \hat{\sigma}_{-}^{(i)}\right).
\end{equation}

We see that the dipole-dipole potential increases linearly with the number of atoms, as seen from the inset of Fig. \ref{fig:Potential_plot}. Although such scenario can greatly improve the effects of dipole-dipole interactions due to its collective nature, one would have to carefully consider near-field interactions among atoms on the same side of the lens. Such scenario adds a complexity to the problem that is beyond the scope of this work, but which could be addressed with the presented mathematical formalism. Furthermore, a far-detuned atom trap based on dipole-dipole interactions could be in principle compatible with other near-resonance cooling techniques (as it can be inferred from Fig. \ref{Fig:rho_11}), creating long-living optical traps.

\section{Summary and Outlook}
\label{Section: Summary}

In this work we have shown that resonant dipole-dipole interactions between two atoms can be amplified in the presence of a lens. Deriving a master equation for the dynamics of two driven atomic dipoles placed near the foci of an aplanatic lens system, we evaluate the modified dispersive and dissipative interactions  between the dipoles, demonstrating a lensing effect in their coupling (Fig.\,\ref{Fig:GammaJ}). We also illustrate that the dipole-dipole coupling increases with an increasing numerical aperture of the lens, and analyze the dependence of the lens mediated dipole-dipole interaction on the atomic polarization. Such modified dipole-dipole interactions can be used, for example, to create a mutual trapping potential for atoms. We demonstrate such a trap potential for the case of an atom interacting with a weakly driven atom via an aplanatic lens (Fig.\,\ref{fig:Potential_plot}),  estimating  the limitations to the trap lifetime due to recoil heating, which could be mitigated by the collectively enhanced potential of $N$ weakly driven atoms.

The present results open a new avenue for engineering long-ranged dipole-dipole interactions in quantum optical systems, facilitating strong resonant dipole-dipole coupling while avoiding the detrimental near-field effects common in nanophotonics platforms. %The proposed scheme leverages imaging methods in quantum optics that allow one to address and manipulate single atoms, which can be applied to a many body quantum optical system.
Lens-mediated dipole-dipole interaction could allow for self-organization of remote atomic systems, where external driving fields can control the dipoles oscillation phases to tailor their amplitude and mutual correlations. The general description of an aplanatic lens presented here encompasses the case of light propagation through a long optical fiber coupled to a lens on each end, a suitable platform for long distance dipole-dipole interfacing. One can consider an extension of the present scheme to a  network of lenses and atoms where one can tailor collective multi-atom interactions in long-ranged systems with lenses. Such lens-modified collective dipole-dipole interactions would increase the versatility and modularity  of quantum optical systems. 
% s an extension of the present analysis, one can apply the same methods to the case of dielectric nano- and micro-spheres, which are relevant systems for exploring  macroscopic quantum phenomena. 

\section*{Acknowledgments} This work was supported by
CONICYT-PAI 77190033 and FONDECYT 11200192 from Chile.

\appendix

\section{Medium-assisted EM field}
\label{App:MQED}
Using the macroscopic QED formalism \cite{Gruner96, BuhmannRev, Buhmann1, Buhmann2}, the Hamiltonian for the vacuum EM field in the presence of the surface can be written as
\eqn{\label{hv}
H_F =\sum_{ \lambda = e,m}\int \dd^3 r \int \dd\omega\,\hbar\omega\, \hat{\mb{f}}^\dagger_\lambda\bkt{\mb{r}, \omega}\cdot\hat{\mb{f}}_\lambda\bkt{\mb{r}, \omega},
}
with $\hat{\mb{f}}^\dagger_\lambda\bkt{\mb{r}, \omega}$ and ${\hat{\mb{f}}_\lambda\bkt{\mb{r}, \omega}}$ as the bosonic creation and annihilation operators respectively that take into account the presence of the media. These are the ladder operators corresponding to the noise polarization ($\lambda = e$) and magnetization  ($\lambda = m$) excitations in the medium-assisted EM field, at frequency $\omega$, created or annihilated at position $\mb{r}$. The medium-assisted bosonic operators obey the canonical commutation relations \eqn{\sbkt{ \hat{\mb{f}}_{\lambda}\bkt{\mb{r}, \omega}, \hat{\mb{f}}_{\lambda'}\bkt{\mb{r}', \omega'} } = \sbkt{ \hat{\mb{f}}^{\dagger}_{\lambda}\bkt{\mb{r}, \omega}, \hat{\mb{f}}^\dagger_{\lambda'}\bkt{\mb{r}', \omega'} }=0,\\
\sbkt{ \hat{\mb{f}}_{\lambda}\bkt{\mb{r}, \omega}, \hat{\mb{f}}^\dagger_{\lambda'}\bkt{\mb{r}', \omega'} } = \delta_{\lambda\lambda'}\delta\bkt{\mb{r} - \mb{r}'}\delta\bkt{\omega - \omega'}.}

The electric  field operator evaluated  at position $\mb{r}_0$ is given as \eqn{\label{Era} &\hat{\mb{E}}\bkt{\mb{r}_0}=\non\\
& \sum_{ \lambda = e,m}\int \dd^3 r\int\dd\omega \sbkt{\dbar {G}_\lambda \bkt{\mb{r}_0, \mb{r}, \omega}\cdot \hat{\mb{f}}_{\lambda}\bkt{\mb{r},  \omega} + \text{H.c.}}.} 
The coefficients $\dbar{G}_\lambda\bkt{\mb{r},\mb{r}',\omega}$  are defined as 
\eqn{\dbar{G}_e \bkt{\mb{r},\mb{r}',\omega}=& i\frac{\omega^2}{c^2} \sqrt{\frac{\hbar}{\pi\epsilon_0}\im[\epsilon \bkt{\mb{r}',\omega}]} \dbar{G}\bkt{\mb{r},\mb{r}',\omega}, \\
\dbar{G}_m \bkt{\mb{r},\mb{r}',\omega}=& i\frac{\omega^2}{c^2} \sqrt{\frac{\hbar}{\pi\epsilon_0}\frac{\im [\mu \bkt{\mb{r}', \omega}]}{\abs{\mu\bkt{\mb{r}',\omega}}^2}}\nabla\times \dbar{G}\bkt{\mb{r},\mb{r}',\omega},}
with $\epsilon(\mb{r},\omega)$ and $\mu(\mb{r},\omega)$ as the space-dependent  permittivity and permeability, and $\dbar{G}\bkt{\mb{r}_1,\mb{r}_2,\omega}$ as the field propagator  near the given boundary conditions \cite{Buhmann1, Buhmann2}.

\section{Derivation of the master equation}
\label{App:ME}

We describe the dynamics of the atomic internal degrees of freedom in terms of a Born-Markov master equation  as follows  \cite{BPbook}:
\eqn{
\label{ME}
&\der{\rho_A}{t} =\non\\
&-\frac{1}{\hbar^{2}}\operatorname{Tr}_{F} \int_{0}^{\infty} \mathrm{d} \tau\sbkt{\tilde H_{A F}(t),\sbkt{\tilde H_{A F}(t-\tau), {\rho}_{A} \otimes\ket{0}\bra{0}}},
}
where $ \rho_A $ corresponds to the density matrix of the two atoms and the EM field is considered to be in a vacuum state.  $\tilde H_{AF}\equiv e^{-i { H_F}t/\hbar } H_{AF}  e^{i { H_F}t/\hbar }$ refers to the interaction Hamiltonian in the interaction picture with respect to the  free  Hamiltonian of the EM field. We have assumed that the atoms and the field are weakly coupled and that the field correlations decay much faster compared to the atomic relaxation time scales \cite{BPbook}.

The above equation can be simplified by separating the RHS into four parts as follows:

\begin{widetext}
\begin{align}
    \frac{\mathrm{d} {\rho}_{A}}{\mathrm{d} t}=& -\underbrace{\frac{1}{\hbar^{2}} \operatorname{Tr}_{F} \int_{0}^{\infty} \mathrm{d} \tau \tilde H_{A F}(t)\tilde H_{A F}(t-\tau)  {\rho}_{A} \otimes\ket{0}\bra{0}}_{(\mathrm{I})} -\underbrace{\frac{1}{\hbar^{2}} \operatorname{Tr}_{F} \int_{0}^{\infty} \mathrm{d} \tau {\rho}_{A} \otimes\ket{0}\bra{0} \tilde H_{A F}(t-\tau) \tilde H_{A F}(t)}_{(\mathrm{II})} \nonumber \\
    &+\underbrace{\frac{1}{\hbar^{2}} \operatorname{Tr}_{F} \int_{0}^{\infty} \mathrm{d} \tau \tilde H_{A F}(t) {\rho}_{A} \otimes\ket{0}\bra{0} \tilde H_{A F}(t-\tau)}_{(\mathrm{III})}+\underbrace{\frac{1}{\hbar^{2}} \operatorname{Tr}_{F} \int_{0}^{\infty} \mathrm{d} \tau \tilde H_{A F}(t-\tau) {\rho}_{A} \otimes\ket{0}\bra{0} \tilde H_{A F}(t)}_{(\mathrm{IV})} \label{Born-Markov app Parts}
\end{align}

We now consider the terms in the above master equation one by one as follows

\eqn{\mr{ (I)} =& -\frac{1}{\hbar^2} \tr_F \int_0^\infty \dd\tau \, \tilde H_{AF}(t)\tilde  H_{AF}(t - \tau)\rho_A\otimes\ket{0}\bra{0}\\
 =& -\frac{1}{\hbar^2} \tr_F \int_0^\infty \dd\tau \,\left[\sum_{i=1,2}\sum_{\lambda} \int \dd^3 r\int\dd\omega\,\cbkt{{\bf d}^\dagger\hat{\sigma}_+^{(i)}e^{-i\bkt{\omega - \omega_D}t}+{\bf d}\hat{\sigma}_-^{(i)}e^{-i\bkt{\omega+ \omega_D}t}}\right.\non\\
&\left.\cdot \dbar{G}_{\lambda}\bkt{{\bf r}_i, {\bf r}, \omega} \cdot \hat{\bf  f}_{\lambda}\bkt{{\bf r}, \omega}+ \hat{\bf  f}^\dagger_{\lambda}\bkt{{\bf r}, \omega, \hat{\bf k}}\cdot\dbar{G}^\dagger_{\lambda}\bkt{{\bf r}_i, {\bf r}, \omega}\cdot\cbkt{{\bf d}^\dagger\hat{\sigma}_+^{(i)}e^{i\bkt{\omega + \omega_D}t}+{\bf d}\hat{\sigma}_-^{(i)}e^{i\bkt{\omega - \omega_D}t}}\right]\non\\
 &\left[\sum_{j=1,2}\sum_{\lambda'} \int \dd^3 r'\int\dd\omega'\,\cbkt{{\bf d}^\dagger\hat{\sigma}_+^{(j)}e^{-i\bkt{\omega' - \omega_D}\bkt{t-\tau}}+{\bf d}\hat{\sigma}_-^{(j)}e^{-i\bkt{\omega' + \omega_D}\bkt{t-\tau}}}\cdot \dbar{G}_{\lambda'}\bkt{{\bf r}_j, {\bf r}', \omega'} \cdot \hat{\bf  f}_{\lambda'}\bkt{{\bf r}', \omega'}\right.\non\\
&\left.+ \hat{\bf  f}^\dagger_{\lambda'}\bkt{{\bf r}', \omega'}\cdot\dbar{G}^\dagger_{\lambda'}\bkt{{\bf r}_j, {\bf r}', \omega'}\cdot\cbkt{{\bf d}^\dagger\hat{\sigma}_+^{(j)}e^{i\bkt{\omega' + \omega_D}\bkt{t-\tau}}+{\bf d}\hat{\sigma}_-^{(j)}e^{i\bkt{\omega' - \omega_D}\bkt{t-\tau}}}\right]\rho_A\otimes\ket{0}\bra{0},
}
where we have used Eq.\,\eqref{Era} in Eq.\,\eqref{Eq:Hint} to express the atom-field interaction Hamiltonian in terms of the medium-assisted bosonic operators. Now taking the trace over the EM field we obtain, 
\eqn{
\bkt{\mr{I}}=& -\frac{1}{\hbar^2}  \int_0^\infty \dd\tau \,\sum_{i,j = 1,2}\sum_{\lambda} \int \dd^3 r\int\dd\omega\,\cbkt{{\bf d}^\dagger\hat{\sigma}_+^{(i)}e^{-i\bkt{\omega - \omega_D}t}+{\bf d}\hat{\sigma}_-^{(i)}e^{-i\bkt{\omega + \omega_D}t}}\cdot \dbar{G}_{\lambda}\bkt{{\bf r}_i, {\bf r}, \omega}\cdot\dbar{G}^\dagger_{\lambda}\bkt{{\bf r}_j, {\bf r}, \omega}\non\\
&\cdot\cbkt{{\bf d}^\dagger\hat{\sigma}_+^{(j)}e^{i\bkt{\omega + \omega_D}\bkt{t-\tau}}+{\bf d}\hat{\sigma}_-^{(j)}e^{i\bkt{\omega - \omega_D}\bkt{t-\tau}}}\rho_A\non\\
=& -\frac{\mu_0 }{\pi\hbar}\int_0^\infty\dd\tau\,\sum_{i,j= 1,2 }\int\dd\omega\,\omega^2\sbkt{{\bf d}^\dagger\hat{\sigma}_+^{(i)}e^{-i\bkt{\omega - \omega_D}t}+{\bf d}\hat{\sigma}_-^{(i)}e^{-i\bkt{\omega + \omega_D}t}}\cdot \im\sbkt{\dbar{G}\bkt{{\bf r}_i, {\bf r}_j, \omega}}\non\\
&\cdot\sbkt{{\bf d}^\dagger\hat{\sigma}_+^{(j)}e^{i\bkt{\omega + \omega_D}\bkt{t-\tau}}+{\bf d}\hat{\sigma}_-^{(j)}e^{i\bkt{\omega - \omega_D}\bkt{t-\tau}}}\rho_A
}
where we have used the relation  $\sum_\lambda \int \dd^3r \, \dbar{G}_\lambda\bkt{{\bf r}_1, {\bf r}, \omega}\cdot \dbar{G}^\dagger_\lambda\bkt{{\bf r}_2, {\bf r}, \omega} = \frac{\hbar\mu_0\omega^2 }{\pi}\im\, \dbar{G}\bkt{{\bf r}_1, {\bf r}_2, \omega}$ for the Green's tensor \cite{Buhmann1}. This can be further simplified in the Markovian limit by performing the time integral using $\int_0^\infty\mathrm{d}\tau e^{i\omega \tau} = \pi\delta(\omega)+i\mathcal{P}\left(\frac{1}{\omega}\right)$, such that the real and imaginary parts are related via the Kramers-Kronig relations. This yields:

\eqn{
\bkt{\mr{I}}=& -\frac{\mu_0 }{\pi\hbar} \sum_{i,j = 1,2}\int\dd\omega\,\omega^2\left[\bkt{{\bf d}^\dagger\cdot \im\sbkt{\dbar{G}\bkt{{\bf r}_i, {\bf r}_j, \omega}}\cdot{\bf d}}\cbkt{\pi\delta \bkt{\omega- \omega_D} -i\mc{P} \frac{1}{\omega - \omega_D}}\hat{\sigma}_+^{(i)}\hat{\sigma}_-^{(j)}\right.\non\\
&\left.+\bkt{{\bf d}^\dagger \cdot \im\sbkt{\dbar{G}\bkt{{\bf r}_i, {\bf r}_j, \omega}}\cdot{\bf d}}\cbkt{\pi\delta \bkt{\omega+ \omega_D} -i\mc{P} \frac{1}{\omega + \omega_D}}\hat{\sigma}_-^{(i)}\hat{\sigma}_+^{(j)}\right]\rho_A\\
 =&\sum_{i,j = 1,2} \bkt{-\frac{\Gamma_{ij}}{2} - \frac{i}{\hbar} J_{ij}^{(+)} } \hat{\sigma}_+^{(i)}\hat{\sigma}_-^{(j)}\rho_A - \frac{i}{\hbar}  J_{ij}^{(-)} \hat{\sigma}_-^{(i)}\hat{\sigma}_+^{(j)}\rho_A
 \label{MEI}
}
where we have defined the quantities $\Gamma_{ij}$, $ J_{ij}^{(+)}$, and $ J_{ij}^{(-)}$ as in Eqs.\,\eqref{Eq:JOR}--\eqref{Eq:gammaij} corresponding to the modification to the collective spontaneous emission and the level shifts respectively. 

Similarly, simplifying the other terms yields:

\eqn{
 \label{MEII}
\mr{ (II)} =& -\frac{1}{\hbar^2} \tr_F \int_0^\infty \dd\tau \,\rho_A\otimes\ket{0}\bra{0} \tilde H_{AF}(t - \tau) \tilde  H_{AF}(t)\non\\
=& \sum_{i,j}\bkt{-\frac{\Gamma_{ij}}{2} + \frac{i}{\hbar} J_{ij}^{(+)} } \rho_A\hat{\sigma}_+^{(i)}\hat{\sigma}_-^{(j)} + \frac{i}{\hbar} J_{ij}^{(-)}\rho_A \hat{\sigma}_-^{(i)}\hat{\sigma}_+^{(j)}\\
\label{MEIII}
 \mr{ (III)} =& \frac{1}{\hbar^2} \tr_F \int_0^\infty \dd\tau \,\tilde H_{AF}(t)\rho_A\otimes\ket{0}\bra{0}\tilde  H_{AF}(t - \tau)\non\\
 =& \sum_{i,j = 1,2}  - \frac{i}{\hbar} J_{ij}^{(+)}  \hat{\sigma}_-^{(i)}\rho_A\hat{\sigma}_+^{(j)} - \frac{i}{\hbar}  J_{ij}^{(-)} \hat{\sigma}_+^{(i)}\rho_A\hat{\sigma}_-^{(j)}+ \frac{1}{2}\sum_l\int d^{3} \mathbf{k}\hat{ \mc{O}}^{(i)}_{\mb{k}l} \hat{\sigma}^{(i)}_- \rho_A \hat{\sigma}^{(j)}_+\bkt{\hat{\mc{O}}^{(j)}_{\mb{k}l}}^\dagger\\
 \label{MEIV}
 \mr{ (IV)} =& \frac{1}{\hbar^2} \tr_F \int_0^\infty \dd\tau \,\tilde  H_{AF}(t - \tau) \rho_A\otimes\ket{0}\bra{0}\tilde H_{AF}(t)\non\\
 =&\sum_{i,j = 1,2}  \frac{i}{\hbar} J_{ij}^{(+)}  \hat{\sigma}_-^{(i)}\rho_A\hat{\sigma}_+^{(j)} + \frac{i}{\hbar}  J_{ij}^{(-)} \hat{\sigma}_+^{(i)}\rho_A\hat{\sigma}_-^{(j)}+ \frac{1}{2} \sum_l\int d^{3} \mathbf{k}\hat{ \mc{O}}^{(i)}_{\mb{k}l} \hat{\sigma}^{(i)}_- \rho_A \hat{\sigma}^{(j)}_+\bkt{\hat{\mc{O}}^{(j)}_{\mb{k}l}}^\dagger,
}
where the jump operators are as defined in Eq.\,\eqref{Eq:jump}.

Thus substituting Eqs.\,\eqref{MEI}, \eqref{MEII}, \eqref{MEIII} and \eqref{MEIV} in Eq.\,\eqref{ME}, we obtain the collective atomic master equation Eq.\,\eqref{Eq:ME}.
\end{widetext}

\section{Derivation of the Green's Tensor near an aplanatic Lens}
\label{App: Green Tensor of a Lens}

\begin{figure}[h]
    \centering
    \includegraphics[width=0.9\linewidth]{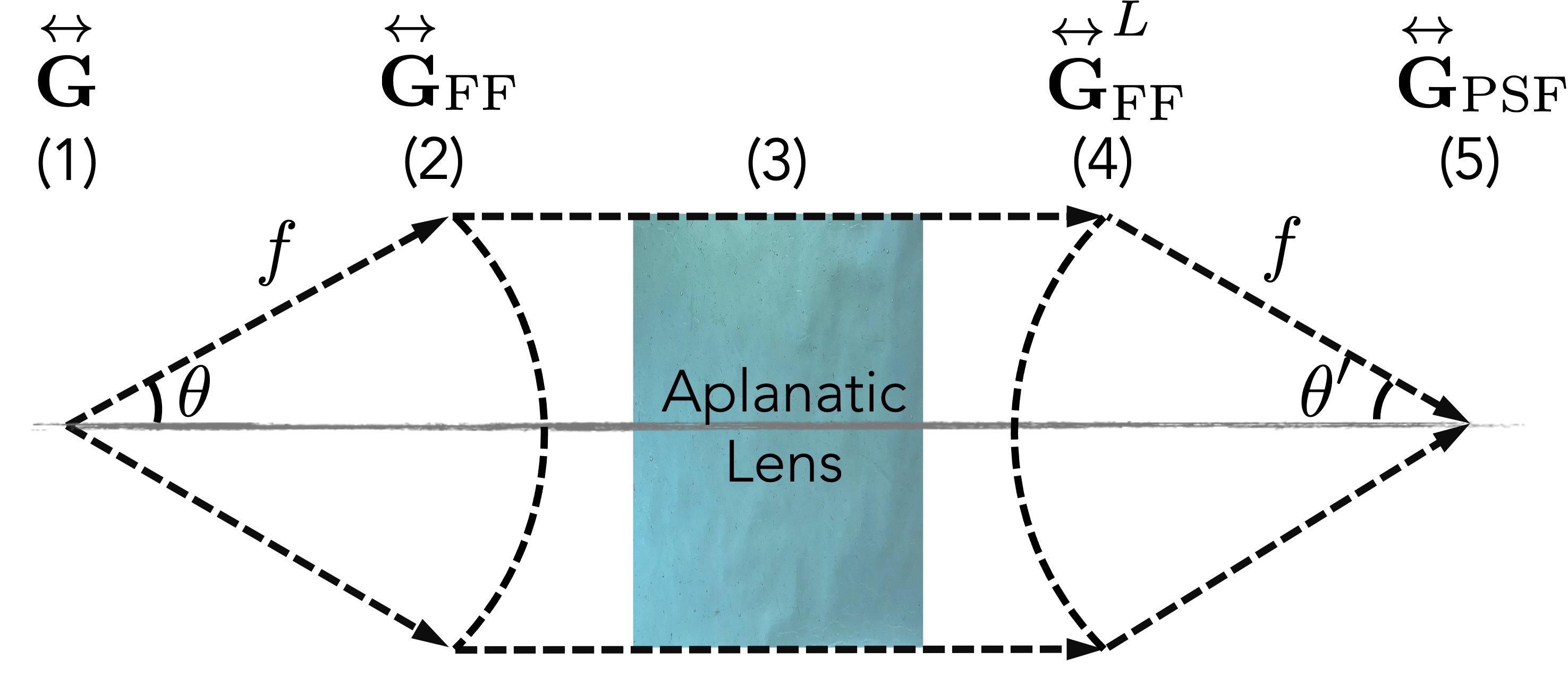}
    \caption{Schematic of the step-by-step propagation of the EM field. (1) A point source dipole radiates from $z_0$ near the left focal region, as describe by the Green's tensor $\dbar{G}$. (2) The far field propagates to the vacuum-lens interface, as described by $\dbar{G}_{\mr{FF}}$. (3) The field propagates through the aplanatic lens as an ideal unperturbed plane wavefront. (4) The lens-vacuum interfase changes the field wavefront to revert the field propagations, as described by $\dbar{G}_{\mr{FF}}^L$. (5) The field is focused down at the focal region, as described by $\dbar{G}_{\mr{PSF}}$.}
    \label{fig:SchematicSteps}
\end{figure}

Our goal is to obtain an expression for the field emitted by a point source upon propagation through an aplanatic lens with equal focal lengths on each side. To do so, we ought to find a \textit{dyadic point-spread function}, as the Green's function in Eq.\,\eqref{FocalField}.
We calculate the field step-by-step as it propagates trough the optical system, aided by its representation in Fig. \ref{fig:SchematicSteps}. It is convenient to analyse the field wavefront as two spherical fronts of radius $f$ centered at the focal point on each side of the lens. The field emitted by a dipole (Fig. \ref{fig:SchematicSteps} (1)) is characterized by its full Green's function in free-space \cite{Novotnybook} 
\begin{align}
    \stackrel{\leftrightarrow}{\mathbf{G}}\left(\mathbf{r}, \mathbf{r}_{0}\right)=\frac{\exp (\mathrm{i} k R)}{4 \pi R} &\left[\left(1+\frac{\mathrm{i} k R-1}{k^{2} R^{2}}\right) \stackrel{\leftrightarrow}{\mathbf{I}}\right.\nonumber\\
    &+\left.\frac{3-3 \mathrm{i} k R-k^{2} R^{2}}{k^{2} R^{2}} \frac{\mathbf{R R}}{R^{2}}\right]\label{Full Green function}, 
\end{align}
where $\mathbf{R}=\mathbf{r}-\mathbf{r}_{0}$, $R=|\mathbf{R}|$ and $\mathbf{R}\mathbf{R}$ denotes the outer product of $\mathbf{R}$ with itself. The emission can be separated into three contributions: the near-field  $\left(G_{\mathrm{NF}}\right)$, intermediate-field $\left(G_{\mathrm{IF}}\right)$, and far-field $\left(G_{\mathrm{FF}}\right)$ Green's tensors,
\begin{align}
\stackrel{\leftrightarrow}{\mathbf{G}}_{\mathrm{NF}}&=\frac{\exp (\mathrm{i} k R)}{4 \pi R} \frac{1}{k^{2} R^{2}}\left[-\overleftrightarrow{\mathbf{I}}+3 \mathbf{R R} / R^{2}\right] \\ 
\stackrel{\leftrightarrow}{\mathbf{G}}_{\mathrm{IF}}&=\frac{\exp (\mathrm{i} k R)}{4 \pi R} \frac{\mathrm{i}}{k R}\left[\overleftrightarrow{\mathbf{I}}-3 \mathbf{R R} / R^{2}\right] \\
\stackrel{\leftrightarrow}{\mathbf{G}}_{\mathrm{FF}}&=\frac{\exp (\mathrm{i} k R)}{4 \pi R}\left[\overleftrightarrow{\mathbf{I}}-\mathbf{R} \mathbf{R} / R^{2}\right].
\end{align}
Since $f\gg\lambda_{0}$, at the lens (Fig. \ref{fig:SchematicSteps}(2)) we are only interested in the far-field Green's function $\left(G_{\mathrm{FF}}\right)$, which can be rewritten in spherical coordinates as
\begin{widetext}
\begin{equation}\label{GreenFunctionFarFieldOrigin}
    \stackrel{\leftrightarrow}{\mathbf{G}}_{\mathrm{FF}}(\mathbf{r}, 0)=\frac{\exp (\mathrm{i} k r)}{4 \pi r}  \left[\begin{array}{lll}
    1-\cos ^{2} \phi \sin ^{2} \theta & -\sin \phi \cos \phi \sin ^{2} \theta & -\cos \phi \sin \theta \cos \theta \\
    -\sin \phi \cos \phi \sin ^{2} \theta & 1-\sin ^{2} \phi \sin ^{2} \theta & -\sin \phi \sin \theta \cos \theta \\
    -\cos \phi \sin \theta \cos \theta & -\sin \phi \sin \theta \cos \theta & \sin ^{2} \theta
    \end{array}\right],
\end{equation}
\end{widetext}
Assuming that the reflection indices of the components of the optical system are negligible, we can think that all the radiative content is collimated between the two reference spheres (Fig. \ref{fig:SchematicSteps} (3)), travelling as plane waves. Just after the second reference sphere, and for a non-reflective aplanatic lens, it is possible to write the output field exactly as the far-field component Eq.(\ref{GreenFunctionFarFieldOrigin}) with opposite sign regarding its original direction of propagation, meaning focusing down instead of diverging away. Eq.\,\eqref{Far Field Lens} shows the field $\mathbf{E}_{\mathrm{FF}}$ just after the linear optical system (Fig. \ref{fig:SchematicSteps} (4)).

In order to obtain a semi-analytic expression for field near the focus (Fig. \ref{fig:SchematicSteps} (5)), we can use the \textit{angular spectrum representation}, Eq.\,\eqref{AngularSpectrumRepresentation}, which allows one to understand the focal field in terms of a series expansion of plane waves with variable amplitudes and propagation directions. The field strength depends on the maximum opening angle $\theta_{\mathrm{max}}$ of the imaging system, and is given by the numerical aperture $\mathrm{NA} = n_{\mathrm{vacuum}} \sin \theta_{\mathrm{max}}$. We assume a homogeneous environment so the refractive index outside the lens can be set to $n\approx 1$. Replacing the far-field towards the focus Eq.\,\eqref{Far Field Lens} into the angular spectrum representation  Eq.\,\eqref{AngularSpectrumRepresentation} gives us $\mathbf{E}_{\mathrm{fo}}$ in terms of the \textit{point-spread Green's function} $\stackrel{\leftrightarrow}{\mathbf{G}}_{\mathrm{PSF}}$. The integrals over the azimuthal angle $\phi$ can be performed analytically using the identities \cite{Hohenesterbook}
\begin{align}\label{Bessel Relations}
    \int_{0}^{2 \pi}\left\{\begin{array}{c}
    \sin n \phi \\
    \cos n \phi
    \end{array}\right\} \mr{e}^{\mathrm{i} x \cos (\phi-\varphi)} \mathrm{d} \phi=2 \pi \mathrm{i}^{n} J_{n}(x)\left\{\begin{array}{c}
    \sin n \varphi \\
    \cos n \varphi
    \end{array}\right\},
\end{align}
where $J_{n}$ are the Bessel functions of order $n$ and $\varphi$ is the azimuthal coordinate for the focal zone, where we will use the cylindrical system $\mb{r}=\left\{\rho,\varphi,z\right\}$, as mentioned before Eq.\eqref{AngularSpectrumRepresentation}. The remaining integral over the polar angle $\theta$, of the form
\begin{equation}\label{PhasesArgIntegrals}
    \int_{0}^{\theta_{\mr{max}}}J_n(k \rho \sin \theta)\left\{\begin{array}{c}
    \sin n \varphi \\
    \cos n \varphi
    \end{array}\right\} \mathrm{e}^{\mathrm{i} k z \cos \theta} \mathrm{e}^{\mathrm{i} k \rho \sin \theta \cos (\phi-\varphi)} \mr{d}\theta,
\end{equation}
does not have a trivial analytic expression and therefore its value must be found for each coordinate $z$ and $\rho$ numerically. The calculation can be generalized for an arbitrary position $\mathbf{r}_0\neq \mathbf{0}$ of the emitter near the focal point of the lens, replacing $r$ by $|\mathbf{r}-\mathbf{r}_0|\approx r-\left(x_{0} x / r+y_{0} y / r+z_{0} z / r\right)=r-(x_0\cos{\phi}+y_0\sin{\phi}+z_0\cos\theta)$. One of the arguments in \eqref{PhasesArgIntegrals} can be rewriten as $\rho \cos(\phi-\varphi)= x_{\rho} \cos \phi + y_{\rho} \sin \phi$,
where $x_{\rho}=\rho\cos\varphi$ and $y_{\rho}=\rho\sin\varphi$. Thus, the phase along the plane transverse to the optical axis can be written as
$\left[ x_{\rho} \cos \phi + y_{\rho} \sin \phi \right] - \left[ x_{0} \cos\phi +y_{0} \sin\phi \right]= \rho_{\text{eff}} \ \cos(\phi - \varphi_{\text{eff}})$, such that the complex exponentials in Eq.\,\eqref{PhasesArgIntegrals} becomes
\begin{align}\label{PhasesArgIntegrals_effective}
   \int_{0}^{\theta_{\mr{max}}}&J_n(k \rho_{\text{eff}} \sin\theta)\left\{\begin{array}{c}
    \sin n \varphi \\
    \cos n \varphi
    \end{array}\right\}\times \nonumber \\
    &\times\mathrm{e}^{\mathrm{i}k z_{\text{eff}}\cos\theta } \mathrm{e}^{\mathrm{i} k \rho_{\text{eff}} \sin\theta   \ \cos(\phi - \varphi_{\text{eff}})} \mr{d}\theta
\end{align}
with the effective coordinates
\begin{align}
    z_{\text{eff}}&=z-z_{0} \\ 
    \rho_{\text{eff}}&= \sqrt{(x_\rho - x_0)^2 + (y_\rho - y_0)^2} \\
    \varphi_{\text{eff}}&= \left\{ \begin{array}{cc}
    \tan^{-1}\frac{y_\rho - y_0}{x_\rho - x_0} & x_\rho - x_0 >0 \\
    \tan^{-1}\frac{y_\rho - y_0}{x_\rho - x_0} + \pi & x_\rho - x_0<0 
    \end{array} \right. .
\end{align}
This allows one to generalize $\left(G_{\mathrm{FF}}\right)$ to source points outside the origin, $\mathbf{r}_{0}\neq 0$, connecting any pair of points between both focal regions and understanding the system in terms of effective coordinates. If the emitter and receiver are in the focal region, then only the relative distance between the two points matters, and not their absolute positions.

One last consideration comes from the study of exchanging the position of the emitter and receiver, which changes both the relative distance $z_{\text{eff}}\rightarrow -z_{\text{eff}}$ and the direction of propagation $k_{z}\rightarrow -k_{z}$. This shows that under exchange of atoms there is no alteration in the arguments of the integral. This is equivalent to having the absolute value $|z_{\text{eff}}|$ in the argument of Eq.\,\eqref{PhasesArgIntegrals_effective}. In this way, we get the dyadic point-spread function Eq.\,\eqref{GreenPSF}, and its matrix elements given by the matrices in Eqs.\,\eqref{Integral_I1}-\eqref{Integral_I4} that represent the propagation of the EM field between any two atoms in either side of the optical system.

\bibliography{Casimir.bib}

\begin{thebibliography}{68}%
\makeatletter
\providecommand \@ifxundefined [1]{%
 \@ifx{#1\undefined}
}%
\providecommand \@ifnum [1]{%
 \ifnum #1\expandafter \@firstoftwo
 \else \expandafter \@secondoftwo
 \fi
}%
\providecommand \@ifx [1]{%
 \ifx #1\expandafter \@firstoftwo
 \else \expandafter \@secondoftwo
 \fi
}%
\providecommand \natexlab [1]{#1}%
\providecommand \enquote  [1]{``#1''}%
\providecommand \bibnamefont  [1]{#1}%
\providecommand \bibfnamefont [1]{#1}%
\providecommand \citenamefont [1]{#1}%
\providecommand \href@noop [0]{\@secondoftwo}%
\providecommand \href [0]{\begingroup \@sanitize@url \@href}%
\providecommand \@href[1]{\@@startlink{#1}\@@href}%
\providecommand \@@href[1]{\endgroup#1\@@endlink}%
\providecommand \@sanitize@url [0]{\catcode `\\12\catcode `\$12\catcode
  `\&12\catcode `\#12\catcode `\^12\catcode `\_12\catcode `\%12\relax}%
\providecommand \@@startlink[1]{}%
\providecommand \@@endlink[0]{}%
\providecommand \url  [0]{\begingroup\@sanitize@url \@url }%
\providecommand \@url [1]{\endgroup\@href {#1}{\urlprefix }}%
\providecommand \urlprefix  [0]{URL }%
\providecommand \Eprint [0]{\href }%
\providecommand \doibase [0]{http://dx.doi.org/}%
\providecommand \selectlanguage [0]{\@gobble}%
\providecommand \bibinfo  [0]{\@secondoftwo}%
\providecommand \bibfield  [0]{\@secondoftwo}%
\providecommand \translation [1]{[#1]}%
\providecommand \BibitemOpen [0]{}%
\providecommand \bibitemStop [0]{}%
\providecommand \bibitemNoStop [0]{.\EOS\space}%
\providecommand \EOS [0]{\spacefactor3000\relax}%
\providecommand \BibitemShut  [1]{\csname bibitem#1\endcsname}%
\let\auto@bib@innerbib\@empty
%</preamble>
\bibitem [{\citenamefont {Bakr}\ \emph {et~al.}(2009)\citenamefont {Bakr},
  \citenamefont {Gillen}, \citenamefont {Peng}, \citenamefont {F{\"o}lling},\
  and\ \citenamefont {Greiner}}]{Bakr2009}%
  \BibitemOpen
  \bibfield  {author} {\bibinfo {author} {\bibfnamefont {W.~S.}\ \bibnamefont
  {Bakr}}, \bibinfo {author} {\bibfnamefont {J.~I.}\ \bibnamefont {Gillen}},
  \bibinfo {author} {\bibfnamefont {A.}~\bibnamefont {Peng}}, \bibinfo {author}
  {\bibfnamefont {S.}~\bibnamefont {F{\"o}lling}}, \ and\ \bibinfo {author}
  {\bibfnamefont {M.}~\bibnamefont {Greiner}},\ }\href {\doibase
  10.1038/nature08482} {\bibfield  {journal} {\bibinfo  {journal} {Nature}\
  }\textbf {\bibinfo {volume} {462}},\ \bibinfo {pages} {74} (\bibinfo {year}
  {2009})}\BibitemShut {NoStop}%
\bibitem [{\citenamefont {Cheuk}\ \emph {et~al.}(2015)\citenamefont {Cheuk},
  \citenamefont {Nichols}, \citenamefont {Okan}, \citenamefont {Gersdorf},
  \citenamefont {Ramasesh}, \citenamefont {Bakr}, \citenamefont {Lompe},\ and\
  \citenamefont {Zwierlein}}]{Cheuk2015}%
  \BibitemOpen
  \bibfield  {author} {\bibinfo {author} {\bibfnamefont {L.~W.}\ \bibnamefont
  {Cheuk}}, \bibinfo {author} {\bibfnamefont {M.~A.}\ \bibnamefont {Nichols}},
  \bibinfo {author} {\bibfnamefont {M.}~\bibnamefont {Okan}}, \bibinfo {author}
  {\bibfnamefont {T.}~\bibnamefont {Gersdorf}}, \bibinfo {author}
  {\bibfnamefont {V.~V.}\ \bibnamefont {Ramasesh}}, \bibinfo {author}
  {\bibfnamefont {W.~S.}\ \bibnamefont {Bakr}}, \bibinfo {author}
  {\bibfnamefont {T.}~\bibnamefont {Lompe}}, \ and\ \bibinfo {author}
  {\bibfnamefont {M.~W.}\ \bibnamefont {Zwierlein}},\ }\href {\doibase
  10.1103/PhysRevLett.114.193001} {\bibfield  {journal} {\bibinfo  {journal}
  {Phys. Rev. Lett.}\ }\textbf {\bibinfo {volume} {114}},\ \bibinfo {pages}
  {193001} (\bibinfo {year} {2015})}\BibitemShut {NoStop}%
\bibitem [{\citenamefont {Parsons}\ \emph {et~al.}(2015)\citenamefont
  {Parsons}, \citenamefont {Huber}, \citenamefont {Mazurenko}, \citenamefont
  {Chiu}, \citenamefont {Setiawan}, \citenamefont {Wooley-Brown}, \citenamefont
  {Blatt},\ and\ \citenamefont {Greiner}}]{Parson2015}%
  \BibitemOpen
  \bibfield  {author} {\bibinfo {author} {\bibfnamefont {M.~F.}\ \bibnamefont
  {Parsons}}, \bibinfo {author} {\bibfnamefont {F.}~\bibnamefont {Huber}},
  \bibinfo {author} {\bibfnamefont {A.}~\bibnamefont {Mazurenko}}, \bibinfo
  {author} {\bibfnamefont {C.~S.}\ \bibnamefont {Chiu}}, \bibinfo {author}
  {\bibfnamefont {W.}~\bibnamefont {Setiawan}}, \bibinfo {author}
  {\bibfnamefont {K.}~\bibnamefont {Wooley-Brown}}, \bibinfo {author}
  {\bibfnamefont {S.}~\bibnamefont {Blatt}}, \ and\ \bibinfo {author}
  {\bibfnamefont {M.}~\bibnamefont {Greiner}},\ }\href {\doibase
  10.1103/PhysRevLett.114.213002} {\bibfield  {journal} {\bibinfo  {journal}
  {Phys. Rev. Lett.}\ }\textbf {\bibinfo {volume} {114}},\ \bibinfo {pages}
  {213002} (\bibinfo {year} {2015})}\BibitemShut {NoStop}%
\bibitem [{\citenamefont {Haller}\ \emph {et~al.}(2015)\citenamefont {Haller},
  \citenamefont {Hudson}, \citenamefont {Kelly}, \citenamefont {Cotta},
  \citenamefont {Peaudecerf}, \citenamefont {Bruce},\ and\ \citenamefont
  {Kuhr}}]{Haller2015}%
  \BibitemOpen
  \bibfield  {author} {\bibinfo {author} {\bibfnamefont {E.}~\bibnamefont
  {Haller}}, \bibinfo {author} {\bibfnamefont {J.}~\bibnamefont {Hudson}},
  \bibinfo {author} {\bibfnamefont {A.}~\bibnamefont {Kelly}}, \bibinfo
  {author} {\bibfnamefont {D.~A.}\ \bibnamefont {Cotta}}, \bibinfo {author}
  {\bibfnamefont {B.}~\bibnamefont {Peaudecerf}}, \bibinfo {author}
  {\bibfnamefont {G.~D.}\ \bibnamefont {Bruce}}, \ and\ \bibinfo {author}
  {\bibfnamefont {S.}~\bibnamefont {Kuhr}},\ }\href {\doibase
  10.1038/nphys3403} {\bibfield  {journal} {\bibinfo  {journal} {Nature
  Physics}\ }\textbf {\bibinfo {volume} {11}},\ \bibinfo {pages} {738}
  (\bibinfo {year} {2015})}\BibitemShut {NoStop}%
\bibitem [{\citenamefont {Yamamoto}\ \emph {et~al.}(2016)\citenamefont
  {Yamamoto}, \citenamefont {Kobayashi}, \citenamefont {Kuno}, \citenamefont
  {Kato},\ and\ \citenamefont {Takahashi}}]{Yamamoto2016}%
  \BibitemOpen
  \bibfield  {author} {\bibinfo {author} {\bibfnamefont {R.}~\bibnamefont
  {Yamamoto}}, \bibinfo {author} {\bibfnamefont {J.}~\bibnamefont {Kobayashi}},
  \bibinfo {author} {\bibfnamefont {T.}~\bibnamefont {Kuno}}, \bibinfo {author}
  {\bibfnamefont {K.}~\bibnamefont {Kato}}, \ and\ \bibinfo {author}
  {\bibfnamefont {Y.}~\bibnamefont {Takahashi}},\ }\href {\doibase
  10.1088/1367-2630/18/2/023016} {\bibfield  {journal} {\bibinfo  {journal}
  {New Journal of Physics}\ }\textbf {\bibinfo {volume} {18}},\ \bibinfo
  {pages} {023016} (\bibinfo {year} {2016})}\BibitemShut {NoStop}%
\bibitem [{\citenamefont {Endres}\ \emph {et~al.}(2016)\citenamefont {Endres},
  \citenamefont {Bernien}, \citenamefont {Keesling}, \citenamefont {Levine},
  \citenamefont {Anschuetz}, \citenamefont {Krajenbrink}, \citenamefont
  {Senko}, \citenamefont {Vuletic}, \citenamefont {Greiner},\ and\
  \citenamefont {Lukin}}]{Endres2016}%
  \BibitemOpen
  \bibfield  {author} {\bibinfo {author} {\bibfnamefont {M.}~\bibnamefont
  {Endres}}, \bibinfo {author} {\bibfnamefont {H.}~\bibnamefont {Bernien}},
  \bibinfo {author} {\bibfnamefont {A.}~\bibnamefont {Keesling}}, \bibinfo
  {author} {\bibfnamefont {H.}~\bibnamefont {Levine}}, \bibinfo {author}
  {\bibfnamefont {E.~R.}\ \bibnamefont {Anschuetz}}, \bibinfo {author}
  {\bibfnamefont {A.}~\bibnamefont {Krajenbrink}}, \bibinfo {author}
  {\bibfnamefont {C.}~\bibnamefont {Senko}}, \bibinfo {author} {\bibfnamefont
  {V.}~\bibnamefont {Vuletic}}, \bibinfo {author} {\bibfnamefont
  {M.}~\bibnamefont {Greiner}}, \ and\ \bibinfo {author} {\bibfnamefont
  {M.~D.}\ \bibnamefont {Lukin}},\ }\href {\doibase 10.1126/science.aah3752}
  {\bibfield  {journal} {\bibinfo  {journal} {Science}\ }\textbf {\bibinfo
  {volume} {354}},\ \bibinfo {pages} {1024} (\bibinfo {year}
  {2016})}\BibitemShut {NoStop}%
\bibitem [{\citenamefont {Barredo}\ \emph {et~al.}(2016)\citenamefont
  {Barredo}, \citenamefont {de~L{\'e}s{\'e}leuc}, \citenamefont {Lienhard},
  \citenamefont {Lahaye},\ and\ \citenamefont {Browaeys}}]{Barredo2016}%
  \BibitemOpen
  \bibfield  {author} {\bibinfo {author} {\bibfnamefont {D.}~\bibnamefont
  {Barredo}}, \bibinfo {author} {\bibfnamefont {S.}~\bibnamefont
  {de~L{\'e}s{\'e}leuc}}, \bibinfo {author} {\bibfnamefont {V.}~\bibnamefont
  {Lienhard}}, \bibinfo {author} {\bibfnamefont {T.}~\bibnamefont {Lahaye}}, \
  and\ \bibinfo {author} {\bibfnamefont {A.}~\bibnamefont {Browaeys}},\ }\href
  {\doibase 10.1126/science.aah3778} {\bibfield  {journal} {\bibinfo  {journal}
  {Science}\ }\textbf {\bibinfo {volume} {354}},\ \bibinfo {pages} {1021}
  (\bibinfo {year} {2016})}\BibitemShut {NoStop}%
\bibitem [{\citenamefont {Kaufman}\ and\ \citenamefont
  {Ni}(2021)}]{Kaufman2021}%
  \BibitemOpen
  \bibfield  {author} {\bibinfo {author} {\bibfnamefont {A.}~\bibnamefont
  {Kaufman}}\ and\ \bibinfo {author} {\bibfnamefont {K.-K.}\ \bibnamefont
  {Ni}},\ }\href {\doibase 10.1038/s41567-021-01357-2} {\bibfield  {journal}
  {\bibinfo  {journal} {Nature Physics}\ }\textbf {\bibinfo {volume} {17}},\
  \bibinfo {pages} {1324} (\bibinfo {year} {2021})}\BibitemShut {NoStop}%
\bibitem [{\citenamefont {Chin}\ \emph {et~al.}(2017)\citenamefont {Chin},
  \citenamefont {Steiner},\ and\ \citenamefont {Kurtsiefer}}]{Chin2017}%
  \BibitemOpen
  \bibfield  {author} {\bibinfo {author} {\bibfnamefont {Y.-S.}\ \bibnamefont
  {Chin}}, \bibinfo {author} {\bibfnamefont {M.}~\bibnamefont {Steiner}}, \
  and\ \bibinfo {author} {\bibfnamefont {C.}~\bibnamefont {Kurtsiefer}},\
  }\href {\doibase 10.1038/s41467-017-01495-3} {\bibfield  {journal} {\bibinfo
  {journal} {Nature Communications}\ }\textbf {\bibinfo {volume} {8}},\
  \bibinfo {pages} {1200} (\bibinfo {year} {2017})}\BibitemShut {NoStop}%
\bibitem [{\citenamefont {Bianchet}\ \emph {et~al.}(2021)\citenamefont
  {Bianchet}, \citenamefont {Alves}, \citenamefont {Zarraoa}, \citenamefont
  {Bruno},\ and\ \citenamefont {Mitchell}}]{Bianchet2021}%
  \BibitemOpen
  \bibfield  {author} {\bibinfo {author} {\bibfnamefont {L.~C.}\ \bibnamefont
  {Bianchet}}, \bibinfo {author} {\bibfnamefont {N.}~\bibnamefont {Alves}},
  \bibinfo {author} {\bibfnamefont {L.}~\bibnamefont {Zarraoa}}, \bibinfo
  {author} {\bibfnamefont {N.}~\bibnamefont {Bruno}}, \ and\ \bibinfo {author}
  {\bibfnamefont {M.~W.}\ \bibnamefont {Mitchell}},\ }\href@noop {} {\enquote
  {\bibinfo {title} {Manipulating and measuring single atoms in the maltese
  cross geometry},}\ } (\bibinfo {year} {2021}),\ \Eprint
  {http://arxiv.org/abs/2103.05783} {arXiv:2103.05783 [physics.atom-ph]}
  \BibitemShut {NoStop}%
\bibitem [{\citenamefont {Robens}\ \emph {et~al.}(2017)\citenamefont {Robens},
  \citenamefont {Brakhane}, \citenamefont {Alt}, \citenamefont {Klei{\ss}ler},
  \citenamefont {Meschede}, \citenamefont {Moon}, \citenamefont {Ramola},\ and\
  \citenamefont {Alberti}}]{Robens17}%
  \BibitemOpen
  \bibfield  {author} {\bibinfo {author} {\bibfnamefont {C.}~\bibnamefont
  {Robens}}, \bibinfo {author} {\bibfnamefont {S.}~\bibnamefont {Brakhane}},
  \bibinfo {author} {\bibfnamefont {W.}~\bibnamefont {Alt}}, \bibinfo {author}
  {\bibfnamefont {F.}~\bibnamefont {Klei{\ss}ler}}, \bibinfo {author}
  {\bibfnamefont {D.}~\bibnamefont {Meschede}}, \bibinfo {author}
  {\bibfnamefont {G.}~\bibnamefont {Moon}}, \bibinfo {author} {\bibfnamefont
  {G.}~\bibnamefont {Ramola}}, \ and\ \bibinfo {author} {\bibfnamefont
  {A.}~\bibnamefont {Alberti}},\ }\href {\doibase 10.1364/OL.42.001043}
  {\bibfield  {journal} {\bibinfo  {journal} {Opt. Lett.}\ }\textbf {\bibinfo
  {volume} {42}},\ \bibinfo {pages} {1043} (\bibinfo {year}
  {2017})}\BibitemShut {NoStop}%
\bibitem [{\citenamefont {Asenjo-Garcia}\ \emph {et~al.}(2017)\citenamefont
  {Asenjo-Garcia}, \citenamefont {Moreno-Cardoner}, \citenamefont {Albrecht},
  \citenamefont {Kimble},\ and\ \citenamefont {Chang}}]{Asenjo17}%
  \BibitemOpen
  \bibfield  {author} {\bibinfo {author} {\bibfnamefont {A.}~\bibnamefont
  {Asenjo-Garcia}}, \bibinfo {author} {\bibfnamefont {M.}~\bibnamefont
  {Moreno-Cardoner}}, \bibinfo {author} {\bibfnamefont {A.}~\bibnamefont
  {Albrecht}}, \bibinfo {author} {\bibfnamefont {H.~J.}\ \bibnamefont
  {Kimble}}, \ and\ \bibinfo {author} {\bibfnamefont {D.~E.}\ \bibnamefont
  {Chang}},\ }\href {\doibase 10.1103/PhysRevX.7.031024} {\bibfield  {journal}
  {\bibinfo  {journal} {Phys. Rev. X}\ }\textbf {\bibinfo {volume} {7}},\
  \bibinfo {pages} {031024} (\bibinfo {year} {2017})}\BibitemShut {NoStop}%
\bibitem [{\citenamefont {Sinha}\ \emph
  {et~al.}(2020{\natexlab{a}})\citenamefont {Sinha}, \citenamefont {Meystre},
  \citenamefont {Goldschmidt}, \citenamefont {Fatemi}, \citenamefont {Rolson},\
  and\ \citenamefont {Solano}}]{Sinha2020}%
  \BibitemOpen
  \bibfield  {author} {\bibinfo {author} {\bibfnamefont {K.}~\bibnamefont
  {Sinha}}, \bibinfo {author} {\bibfnamefont {P.}~\bibnamefont {Meystre}},
  \bibinfo {author} {\bibfnamefont {E.}~\bibnamefont {Goldschmidt}}, \bibinfo
  {author} {\bibfnamefont {F.~K.}\ \bibnamefont {Fatemi}}, \bibinfo {author}
  {\bibfnamefont {S.~L.}\ \bibnamefont {Rolson}}, \ and\ \bibinfo {author}
  {\bibfnamefont {P.}~\bibnamefont {Solano}},\ }\href {\doibase
  10.1103/PhysRevLett.124.043603} {\bibfield  {journal} {\bibinfo  {journal}
  {Phys. Rev. Lett.}\ }\textbf {\bibinfo {volume} {124}},\ \bibinfo {pages}
  {043603} (\bibinfo {year} {2020}{\natexlab{a}})}\BibitemShut {NoStop}%
\bibitem [{\citenamefont {Sinha}\ \emph {et~al.}(2019)\citenamefont {Sinha},
  \citenamefont {Meystre},\ and\ \citenamefont {Solano}}]{Sinha2019}%
  \BibitemOpen
  \bibfield  {author} {\bibinfo {author} {\bibfnamefont {K.}~\bibnamefont
  {Sinha}}, \bibinfo {author} {\bibfnamefont {P.}~\bibnamefont {Meystre}}, \
  and\ \bibinfo {author} {\bibfnamefont {P.}~\bibnamefont {Solano}},\ }\href
  {\doibase 10.1117/12.2530927} {\bibfield  {journal} {\bibinfo  {journal}
  {Nanophotonic Materials, Devices, and Systems}\ }\textbf {\bibinfo {volume}
  {11091}},\ \bibinfo {pages} {53 } (\bibinfo {year} {2019})}\BibitemShut
  {NoStop}%
\bibitem [{\citenamefont {Dinc}\ and\ \citenamefont
  {Bra\`{n}czyk}(2019)}]{Dinc19}%
  \BibitemOpen
  \bibfield  {author} {\bibinfo {author} {\bibfnamefont {F.}~\bibnamefont
  {Dinc}}\ and\ \bibinfo {author} {\bibfnamefont {A.~M.}\ \bibnamefont
  {Bra\`{n}czyk}},\ }\href {\doibase 10.1103/PhysRevResearch.1.032042}
  {\bibfield  {journal} {\bibinfo  {journal} {Phys. Rev. Research}\ }\textbf
  {\bibinfo {volume} {1}},\ \bibinfo {pages} {032042(R)} (\bibinfo {year}
  {2019})}\BibitemShut {NoStop}%
\bibitem [{\citenamefont {Calaj\'{o}}\ \emph {et~al.}(2019)\citenamefont
  {Calaj\'{o}}, \citenamefont {Fang}, \citenamefont {Baranger},\ and\
  \citenamefont {Ciccarello}}]{Calajo19}%
  \BibitemOpen
  \bibfield  {author} {\bibinfo {author} {\bibfnamefont {G.}~\bibnamefont
  {Calaj\'{o}}}, \bibinfo {author} {\bibfnamefont {Y.-L.~L.}\ \bibnamefont
  {Fang}}, \bibinfo {author} {\bibfnamefont {H.~U.}\ \bibnamefont {Baranger}},
  \ and\ \bibinfo {author} {\bibfnamefont {F.}~\bibnamefont {Ciccarello}},\
  }\href {\doibase 10.1103/PhysRevLett.122.073601} {\bibfield  {journal}
  {\bibinfo  {journal} {Phys. Rev. Lett.}\ }\textbf {\bibinfo {volume} {122}},\
  \bibinfo {pages} {073601} (\bibinfo {year} {2019})}\BibitemShut {NoStop}%
\bibitem [{\citenamefont {Sheremet}\ \emph {et~al.}(2021)\citenamefont
  {Sheremet}, \citenamefont {Petrov}, \citenamefont {Iorsh}, \citenamefont
  {Poshakinskiy},\ and\ \citenamefont {Poddubny}}]{Sheremet2021WaveguideQE}%
  \BibitemOpen
  \bibfield  {author} {\bibinfo {author} {\bibfnamefont {A.~S.}\ \bibnamefont
  {Sheremet}}, \bibinfo {author} {\bibfnamefont {M.~I.}\ \bibnamefont
  {Petrov}}, \bibinfo {author} {\bibfnamefont {I.~V.}\ \bibnamefont {Iorsh}},
  \bibinfo {author} {\bibfnamefont {A.~V.}\ \bibnamefont {Poshakinskiy}}, \
  and\ \bibinfo {author} {\bibfnamefont {A.~N.}\ \bibnamefont {Poddubny}}\
  }(\bibinfo {year} {2021})\BibitemShut {NoStop}%
\bibitem [{\citenamefont {Trivedi}\ \emph {et~al.}(2021)\citenamefont
  {Trivedi}, \citenamefont {Malz}, \citenamefont {Sun}, \citenamefont {Fan},\
  and\ \citenamefont {Vu\ifmmode \check{c}\else
  \v{c}\fi{}kovi\ifmmode~\acute{c}\else \'{c}\fi{}}}]{Trivedi21}%
  \BibitemOpen
  \bibfield  {author} {\bibinfo {author} {\bibfnamefont {R.}~\bibnamefont
  {Trivedi}}, \bibinfo {author} {\bibfnamefont {D.}~\bibnamefont {Malz}},
  \bibinfo {author} {\bibfnamefont {S.}~\bibnamefont {Sun}}, \bibinfo {author}
  {\bibfnamefont {S.}~\bibnamefont {Fan}}, \ and\ \bibinfo {author}
  {\bibfnamefont {J.}~\bibnamefont {Vu\ifmmode \check{c}\else
  \v{c}\fi{}kovi\ifmmode~\acute{c}\else \'{c}\fi{}}},\ }\href {\doibase
  10.1103/PhysRevA.104.013705} {\bibfield  {journal} {\bibinfo  {journal}
  {Phys. Rev. A}\ }\textbf {\bibinfo {volume} {104}},\ \bibinfo {pages}
  {013705} (\bibinfo {year} {2021})}\BibitemShut {NoStop}%
\bibitem [{\citenamefont {Buonaiuto}\ \emph {et~al.}(2021)\citenamefont
  {Buonaiuto}, \citenamefont {Carollo}, \citenamefont {Olmos},\ and\
  \citenamefont {Lesanovsky}}]{Buonaiuto21}%
  \BibitemOpen
  \bibfield  {author} {\bibinfo {author} {\bibfnamefont {G.}~\bibnamefont
  {Buonaiuto}}, \bibinfo {author} {\bibfnamefont {F.}~\bibnamefont {Carollo}},
  \bibinfo {author} {\bibfnamefont {B.}~\bibnamefont {Olmos}}, \ and\ \bibinfo
  {author} {\bibfnamefont {I.}~\bibnamefont {Lesanovsky}},\ }\href {\doibase
  10.1103/PhysRevLett.127.133601} {\bibfield  {journal} {\bibinfo  {journal}
  {Phys. Rev. Lett.}\ }\textbf {\bibinfo {volume} {127}},\ \bibinfo {pages}
  {133601} (\bibinfo {year} {2021})}\BibitemShut {NoStop}%
\bibitem [{\citenamefont {Poshakinskiy}\ and\ \citenamefont
  {Poddubny}(2021)}]{Poshakinskiy21}%
  \BibitemOpen
  \bibfield  {author} {\bibinfo {author} {\bibfnamefont {A.~V.}\ \bibnamefont
  {Poshakinskiy}}\ and\ \bibinfo {author} {\bibfnamefont {A.~N.}\ \bibnamefont
  {Poddubny}},\ }\href {\doibase 10.1103/PhysRevLett.127.173601} {\bibfield
  {journal} {\bibinfo  {journal} {Phys. Rev. Lett.}\ }\textbf {\bibinfo
  {volume} {127}},\ \bibinfo {pages} {173601} (\bibinfo {year}
  {2021})}\BibitemShut {NoStop}%
\bibitem [{\citenamefont {Pivovarov}\ \emph {et~al.}(2021)\citenamefont
  {Pivovarov}, \citenamefont {Gerasimov}, \citenamefont {Berroir},
  \citenamefont {Ray}, \citenamefont {Laurat}, \citenamefont {Urvoy},\ and\
  \citenamefont {Kupriyanov}}]{Pivovarov21}%
  \BibitemOpen
  \bibfield  {author} {\bibinfo {author} {\bibfnamefont {V.~A.}\ \bibnamefont
  {Pivovarov}}, \bibinfo {author} {\bibfnamefont {L.~V.}\ \bibnamefont
  {Gerasimov}}, \bibinfo {author} {\bibfnamefont {J.}~\bibnamefont {Berroir}},
  \bibinfo {author} {\bibfnamefont {T.}~\bibnamefont {Ray}}, \bibinfo {author}
  {\bibfnamefont {J.}~\bibnamefont {Laurat}}, \bibinfo {author} {\bibfnamefont
  {A.}~\bibnamefont {Urvoy}}, \ and\ \bibinfo {author} {\bibfnamefont {D.~V.}\
  \bibnamefont {Kupriyanov}},\ }\href {\doibase 10.1103/PhysRevA.103.043716}
  {\bibfield  {journal} {\bibinfo  {journal} {Phys. Rev. A}\ }\textbf {\bibinfo
  {volume} {103}},\ \bibinfo {pages} {043716} (\bibinfo {year}
  {2021})}\BibitemShut {NoStop}%
\bibitem [{\citenamefont {Jones}\ \emph {et~al.}(2020)\citenamefont {Jones},
  \citenamefont {Buonaiuto}, \citenamefont {Lang}, \citenamefont {Lesanovsky},\
  and\ \citenamefont {Olmos}}]{Jones20}%
  \BibitemOpen
  \bibfield  {author} {\bibinfo {author} {\bibfnamefont {R.}~\bibnamefont
  {Jones}}, \bibinfo {author} {\bibfnamefont {G.}~\bibnamefont {Buonaiuto}},
  \bibinfo {author} {\bibfnamefont {B.}~\bibnamefont {Lang}}, \bibinfo {author}
  {\bibfnamefont {I.}~\bibnamefont {Lesanovsky}}, \ and\ \bibinfo {author}
  {\bibfnamefont {B.}~\bibnamefont {Olmos}},\ }\href {\doibase
  10.1103/PhysRevLett.124.093601} {\bibfield  {journal} {\bibinfo  {journal}
  {Phys. Rev. Lett.}\ }\textbf {\bibinfo {volume} {124}},\ \bibinfo {pages}
  {093601} (\bibinfo {year} {2020})}\BibitemShut {NoStop}%
\bibitem [{\citenamefont {Arranz~Regidor}\ \emph {et~al.}(2021)\citenamefont
  {Arranz~Regidor}, \citenamefont {Crowder}, \citenamefont {Carmichael},\ and\
  \citenamefont {Hughes}}]{Hughes2021}%
  \BibitemOpen
  \bibfield  {author} {\bibinfo {author} {\bibfnamefont {S.}~\bibnamefont
  {Arranz~Regidor}}, \bibinfo {author} {\bibfnamefont {G.}~\bibnamefont
  {Crowder}}, \bibinfo {author} {\bibfnamefont {H.}~\bibnamefont {Carmichael}},
  \ and\ \bibinfo {author} {\bibfnamefont {S.}~\bibnamefont {Hughes}},\ }\href
  {\doibase 10.1103/PhysRevResearch.3.023030} {\bibfield  {journal} {\bibinfo
  {journal} {Phys. Rev. Research}\ }\textbf {\bibinfo {volume} {3}},\ \bibinfo
  {pages} {023030} (\bibinfo {year} {2021})}\BibitemShut {NoStop}%
\bibitem [{\citenamefont {van Loo}\ \emph {et~al.}(2013)\citenamefont {van
  Loo}, \citenamefont {Fedorov}, \citenamefont {Lalumi\'{e}re}, \citenamefont
  {Sanders}, \citenamefont {Blais},\ and\ \citenamefont
  {Wallraff}}]{vanLoo2013}%
  \BibitemOpen
  \bibfield  {author} {\bibinfo {author} {\bibfnamefont {A.~F.}\ \bibnamefont
  {van Loo}}, \bibinfo {author} {\bibfnamefont {A.}~\bibnamefont {Fedorov}},
  \bibinfo {author} {\bibfnamefont {K.}~\bibnamefont {Lalumi\'{e}re}}, \bibinfo
  {author} {\bibfnamefont {B.~C.}\ \bibnamefont {Sanders}}, \bibinfo {author}
  {\bibfnamefont {A.}~\bibnamefont {Blais}}, \ and\ \bibinfo {author}
  {\bibfnamefont {A.}~\bibnamefont {Wallraff}},\ }\href {\doibase
  10.1126/science.1244324} {\bibfield  {journal} {\bibinfo  {journal}
  {Science}\ }\textbf {\bibinfo {volume} {342}},\ \bibinfo {pages} {1494}
  (\bibinfo {year} {2013})}\BibitemShut {NoStop}%
\bibitem [{\citenamefont {Solano}\ \emph {et~al.}(2017)\citenamefont {Solano},
  \citenamefont {Barberis-Blostein}, \citenamefont {Fatemi}, \citenamefont
  {Orozco},\ and\ \citenamefont {Rolston}}]{Solano2017}%
  \BibitemOpen
  \bibfield  {author} {\bibinfo {author} {\bibfnamefont {P.}~\bibnamefont
  {Solano}}, \bibinfo {author} {\bibfnamefont {P.}~\bibnamefont
  {Barberis-Blostein}}, \bibinfo {author} {\bibfnamefont {F.~K.}\ \bibnamefont
  {Fatemi}}, \bibinfo {author} {\bibfnamefont {L.~A.}\ \bibnamefont {Orozco}},
  \ and\ \bibinfo {author} {\bibfnamefont {S.~L.}\ \bibnamefont {Rolston}},\
  }\href {\doibase 10.1038/s41467-017-01994-3} {\bibfield  {journal} {\bibinfo
  {journal} {Nat. Commun.}\ }\textbf {\bibinfo {volume} {8}},\ \bibinfo {pages}
  {1857} (\bibinfo {year} {2017})}\BibitemShut {NoStop}%
\bibitem [{\citenamefont {Kim}\ \emph {et~al.}(2018)\citenamefont {Kim},
  \citenamefont {Aghaeimeibodi}, \citenamefont {Richardson}, \citenamefont
  {Leavitt},\ and\ \citenamefont {Waks}}]{Kim2018}%
  \BibitemOpen
  \bibfield  {author} {\bibinfo {author} {\bibfnamefont {J.-H.}\ \bibnamefont
  {Kim}}, \bibinfo {author} {\bibfnamefont {S.}~\bibnamefont {Aghaeimeibodi}},
  \bibinfo {author} {\bibfnamefont {C.~J.~K.}\ \bibnamefont {Richardson}},
  \bibinfo {author} {\bibfnamefont {R.~P.}\ \bibnamefont {Leavitt}}, \ and\
  \bibinfo {author} {\bibfnamefont {E.}~\bibnamefont {Waks}},\ }\href {\doibase
  10.1021/acs.nanolett.8b01133} {\bibfield  {journal} {\bibinfo  {journal}
  {Nano Letters}\ }\textbf {\bibinfo {volume} {18}},\ \bibinfo {pages} {4734}
  (\bibinfo {year} {2018})}\BibitemShut {NoStop}%
\bibitem [{\citenamefont {Newman}\ \emph {et~al.}(2018)\citenamefont {Newman},
  \citenamefont {Cortes}, \citenamefont {Afshar}, \citenamefont {Cadien},
  \citenamefont {Meldrum}, \citenamefont {Fedosejevs},\ and\ \citenamefont
  {Jacob}}]{Newman2018}%
  \BibitemOpen
  \bibfield  {author} {\bibinfo {author} {\bibfnamefont {W.~D.}\ \bibnamefont
  {Newman}}, \bibinfo {author} {\bibfnamefont {C.~L.}\ \bibnamefont {Cortes}},
  \bibinfo {author} {\bibfnamefont {A.}~\bibnamefont {Afshar}}, \bibinfo
  {author} {\bibfnamefont {K.}~\bibnamefont {Cadien}}, \bibinfo {author}
  {\bibfnamefont {A.}~\bibnamefont {Meldrum}}, \bibinfo {author} {\bibfnamefont
  {R.}~\bibnamefont {Fedosejevs}}, \ and\ \bibinfo {author} {\bibfnamefont
  {Z.}~\bibnamefont {Jacob}},\ }\href {\doibase 10.1126/sciadv.aar5278}
  {\bibfield  {journal} {\bibinfo  {journal} {Science Advances}\ }\textbf
  {\bibinfo {volume} {4}},\ \bibinfo {pages} {5278} (\bibinfo {year}
  {2018})}\BibitemShut {NoStop}%
\bibitem [{\citenamefont {Boddeti}\ \emph {et~al.}(2022)\citenamefont
  {Boddeti}, \citenamefont {Guan}, \citenamefont {Sentz}, \citenamefont
  {Juarez}, \citenamefont {Newman}, \citenamefont {Cortes}, \citenamefont
  {Odom},\ and\ \citenamefont {Jacob}}]{Boddeti2022}%
  \BibitemOpen
  \bibfield  {author} {\bibinfo {author} {\bibfnamefont {A.~K.}\ \bibnamefont
  {Boddeti}}, \bibinfo {author} {\bibfnamefont {J.}~\bibnamefont {Guan}},
  \bibinfo {author} {\bibfnamefont {T.}~\bibnamefont {Sentz}}, \bibinfo
  {author} {\bibfnamefont {X.}~\bibnamefont {Juarez}}, \bibinfo {author}
  {\bibfnamefont {W.}~\bibnamefont {Newman}}, \bibinfo {author} {\bibfnamefont
  {C.}~\bibnamefont {Cortes}}, \bibinfo {author} {\bibfnamefont {T.~W.}\
  \bibnamefont {Odom}}, \ and\ \bibinfo {author} {\bibfnamefont
  {Z.}~\bibnamefont {Jacob}},\ }\href {\doibase 10.1021/acs.nanolett.1c02835}
  {\bibfield  {journal} {\bibinfo  {journal} {Nano Letters}\ }\textbf {\bibinfo
  {volume} {22}},\ \bibinfo {pages} {22} (\bibinfo {year} {2022})}\BibitemShut
  {NoStop}%
\bibitem [{\citenamefont {Fermani}\ \emph {et~al.}(2007)\citenamefont
  {Fermani}, \citenamefont {Scheel},\ and\ \citenamefont {Knight}}]{Fermani07}%
  \BibitemOpen
  \bibfield  {author} {\bibinfo {author} {\bibfnamefont {R.}~\bibnamefont
  {Fermani}}, \bibinfo {author} {\bibfnamefont {S.}~\bibnamefont {Scheel}}, \
  and\ \bibinfo {author} {\bibfnamefont {P.~L.}\ \bibnamefont {Knight}},\
  }\href {\doibase 10.1103/PhysRevA.75.062905} {\bibfield  {journal} {\bibinfo
  {journal} {Phys. Rev. A}\ }\textbf {\bibinfo {volume} {75}},\ \bibinfo
  {pages} {062905} (\bibinfo {year} {2007})}\BibitemShut {NoStop}%
\bibitem [{\citenamefont {Yeung}\ and\ \citenamefont
  {Gustafson}(1996)}]{Yeung96}%
  \BibitemOpen
  \bibfield  {author} {\bibinfo {author} {\bibfnamefont {M.~S.}\ \bibnamefont
  {Yeung}}\ and\ \bibinfo {author} {\bibfnamefont {T.~K.}\ \bibnamefont
  {Gustafson}},\ }\href {\doibase 10.1103/PhysRevA.54.5227} {\bibfield
  {journal} {\bibinfo  {journal} {Phys. Rev. A}\ }\textbf {\bibinfo {volume}
  {54}},\ \bibinfo {pages} {5227} (\bibinfo {year} {1996})}\BibitemShut
  {NoStop}%
\bibitem [{\citenamefont {Scheel}\ \emph {et~al.}(2005)\citenamefont {Scheel},
  \citenamefont {Rekdal}, \citenamefont {Knight},\ and\ \citenamefont
  {Hinds}}]{Scheel05}%
  \BibitemOpen
  \bibfield  {author} {\bibinfo {author} {\bibfnamefont {S.}~\bibnamefont
  {Scheel}}, \bibinfo {author} {\bibfnamefont {P.~K.}\ \bibnamefont {Rekdal}},
  \bibinfo {author} {\bibfnamefont {P.~L.}\ \bibnamefont {Knight}}, \ and\
  \bibinfo {author} {\bibfnamefont {E.~A.}\ \bibnamefont {Hinds}},\ }\href
  {\doibase 10.1103/PhysRevA.72.042901} {\bibfield  {journal} {\bibinfo
  {journal} {Phys. Rev. A}\ }\textbf {\bibinfo {volume} {72}},\ \bibinfo
  {pages} {042901} (\bibinfo {year} {2005})}\BibitemShut {NoStop}%
\bibitem [{\citenamefont {Rekdal}\ \emph {et~al.}(2004)\citenamefont {Rekdal},
  \citenamefont {Scheel}, \citenamefont {Knight},\ and\ \citenamefont
  {Hinds}}]{Rekdal04}%
  \BibitemOpen
  \bibfield  {author} {\bibinfo {author} {\bibfnamefont {P.~K.}\ \bibnamefont
  {Rekdal}}, \bibinfo {author} {\bibfnamefont {S.}~\bibnamefont {Scheel}},
  \bibinfo {author} {\bibfnamefont {P.~L.}\ \bibnamefont {Knight}}, \ and\
  \bibinfo {author} {\bibfnamefont {E.~A.}\ \bibnamefont {Hinds}},\ }\href
  {\doibase 10.1103/PhysRevA.70.013811} {\bibfield  {journal} {\bibinfo
  {journal} {Phys. Rev. A}\ }\textbf {\bibinfo {volume} {70}},\ \bibinfo
  {pages} {013811} (\bibinfo {year} {2004})}\BibitemShut {NoStop}%
\bibitem [{\citenamefont {Skagerstam}\ \emph {et~al.}(2006)\citenamefont
  {Skagerstam}, \citenamefont {Hohenester}, \citenamefont {Eiguren},\ and\
  \citenamefont {Rekdal}}]{Skagerstam06}%
  \BibitemOpen
  \bibfield  {author} {\bibinfo {author} {\bibfnamefont {B.-S.~K.}\
  \bibnamefont {Skagerstam}}, \bibinfo {author} {\bibfnamefont
  {U.}~\bibnamefont {Hohenester}}, \bibinfo {author} {\bibfnamefont
  {A.}~\bibnamefont {Eiguren}}, \ and\ \bibinfo {author} {\bibfnamefont
  {P.~K.}\ \bibnamefont {Rekdal}},\ }\href {\doibase
  10.1103/PhysRevLett.97.070401} {\bibfield  {journal} {\bibinfo  {journal}
  {Phys. Rev. Lett.}\ }\textbf {\bibinfo {volume} {97}},\ \bibinfo {pages}
  {070401} (\bibinfo {year} {2006})}\BibitemShut {NoStop}%
\bibitem [{\citenamefont {Sagu\'e}\ \emph {et~al.}(2007)\citenamefont
  {Sagu\'e}, \citenamefont {Vetsch}, \citenamefont {Alt}, \citenamefont
  {Meschede},\ and\ \citenamefont {Rauschenbeutel}}]{Sague07}%
  \BibitemOpen
  \bibfield  {author} {\bibinfo {author} {\bibfnamefont {G.}~\bibnamefont
  {Sagu\'e}}, \bibinfo {author} {\bibfnamefont {E.}~\bibnamefont {Vetsch}},
  \bibinfo {author} {\bibfnamefont {W.}~\bibnamefont {Alt}}, \bibinfo {author}
  {\bibfnamefont {D.}~\bibnamefont {Meschede}}, \ and\ \bibinfo {author}
  {\bibfnamefont {A.}~\bibnamefont {Rauschenbeutel}},\ }\href {\doibase
  10.1103/PhysRevLett.99.163602} {\bibfield  {journal} {\bibinfo  {journal}
  {Phys. Rev. Lett.}\ }\textbf {\bibinfo {volume} {99}},\ \bibinfo {pages}
  {163602} (\bibinfo {year} {2007})}\BibitemShut {NoStop}%
\bibitem [{\citenamefont {Goldstein}\ and\ \citenamefont
  {Meystre}(1997)}]{Goldstein97}%
  \BibitemOpen
  \bibfield  {author} {\bibinfo {author} {\bibfnamefont {E.~V.}\ \bibnamefont
  {Goldstein}}\ and\ \bibinfo {author} {\bibfnamefont {P.}~\bibnamefont
  {Meystre}},\ }\href {\doibase 10.1103/PhysRevA.56.5135} {\bibfield  {journal}
  {\bibinfo  {journal} {Phys. Rev. A}\ }\textbf {\bibinfo {volume} {56}},\
  \bibinfo {pages} {5135} (\bibinfo {year} {1997})}\BibitemShut {NoStop}%
\bibitem [{\citenamefont {Milonni}\ and\ \citenamefont
  {Knight}(1974)}]{Milonni74}%
  \BibitemOpen
  \bibfield  {author} {\bibinfo {author} {\bibfnamefont {P.~W.}\ \bibnamefont
  {Milonni}}\ and\ \bibinfo {author} {\bibfnamefont {P.~L.}\ \bibnamefont
  {Knight}},\ }\href {\doibase 10.1103/PhysRevA.10.1096} {\bibfield  {journal}
  {\bibinfo  {journal} {Phys. Rev. A}\ }\textbf {\bibinfo {volume} {10}},\
  \bibinfo {pages} {1096} (\bibinfo {year} {1974})}\BibitemShut {NoStop}%
\bibitem [{\citenamefont {Sinha}\ \emph
  {et~al.}(2020{\natexlab{b}})\citenamefont {Sinha}, \citenamefont
  {Gonz\'alez-Tudela}, \citenamefont {Lu},\ and\ \citenamefont
  {Solano}}]{Sinha2020b}%
  \BibitemOpen
  \bibfield  {author} {\bibinfo {author} {\bibfnamefont {K.}~\bibnamefont
  {Sinha}}, \bibinfo {author} {\bibfnamefont {A.}~\bibnamefont
  {Gonz\'alez-Tudela}}, \bibinfo {author} {\bibfnamefont {Y.}~\bibnamefont
  {Lu}}, \ and\ \bibinfo {author} {\bibfnamefont {P.}~\bibnamefont {Solano}},\
  }\href {\doibase 10.1103/PhysRevA.102.043718} {\bibfield  {journal} {\bibinfo
   {journal} {Phys. Rev. A}\ }\textbf {\bibinfo {volume} {102}},\ \bibinfo
  {pages} {043718} (\bibinfo {year} {2020}{\natexlab{b}})}\BibitemShut
  {NoStop}%
\bibitem [{\citenamefont {Dung}\ \emph {et~al.}(2002)\citenamefont {Dung},
  \citenamefont {Kn\"oll},\ and\ \citenamefont {Welsch}}]{Dung02}%
  \BibitemOpen
  \bibfield  {author} {\bibinfo {author} {\bibfnamefont {H.~T.}\ \bibnamefont
  {Dung}}, \bibinfo {author} {\bibfnamefont {L.}~\bibnamefont {Kn\"oll}}, \
  and\ \bibinfo {author} {\bibfnamefont {D.-G.}\ \bibnamefont {Welsch}},\
  }\href {\doibase 10.1103/PhysRevA.66.063810} {\bibfield  {journal} {\bibinfo
  {journal} {Phys. Rev. A}\ }\textbf {\bibinfo {volume} {66}},\ \bibinfo
  {pages} {063810} (\bibinfo {year} {2002})}\BibitemShut {NoStop}%
\bibitem [{\citenamefont {Kobayashi}\ \emph {et~al.}(1995)\citenamefont
  {Kobayashi}, \citenamefont {Zheng},\ and\ \citenamefont
  {Sekiguchi}}]{Kobayashi95}%
  \BibitemOpen
  \bibfield  {author} {\bibinfo {author} {\bibfnamefont {T.}~\bibnamefont
  {Kobayashi}}, \bibinfo {author} {\bibfnamefont {Q.}~\bibnamefont {Zheng}}, \
  and\ \bibinfo {author} {\bibfnamefont {T.}~\bibnamefont {Sekiguchi}},\ }\href
  {\doibase 10.1103/PhysRevA.52.2835} {\bibfield  {journal} {\bibinfo
  {journal} {Phys. Rev. A}\ }\textbf {\bibinfo {volume} {52}},\ \bibinfo
  {pages} {2835} (\bibinfo {year} {1995})}\BibitemShut {NoStop}%
\bibitem [{\citenamefont {Goldstein}\ \emph {et~al.}(1996)\citenamefont
  {Goldstein}, \citenamefont {Pax},\ and\ \citenamefont
  {Meystre}}]{Goldstein96}%
  \BibitemOpen
  \bibfield  {author} {\bibinfo {author} {\bibfnamefont {E.~V.}\ \bibnamefont
  {Goldstein}}, \bibinfo {author} {\bibfnamefont {P.}~\bibnamefont {Pax}}, \
  and\ \bibinfo {author} {\bibfnamefont {P.}~\bibnamefont {Meystre}},\ }\href
  {\doibase 10.1103/PhysRevA.53.2604} {\bibfield  {journal} {\bibinfo
  {journal} {Phys. Rev. A}\ }\textbf {\bibinfo {volume} {53}},\ \bibinfo
  {pages} {2604} (\bibinfo {year} {1996})}\BibitemShut {NoStop}%
\bibitem [{\citenamefont {El-Ganainy}\ and\ \citenamefont
  {John}(2013)}]{ElGanainy2013}%
  \BibitemOpen
  \bibfield  {author} {\bibinfo {author} {\bibfnamefont {R.}~\bibnamefont
  {El-Ganainy}}\ and\ \bibinfo {author} {\bibfnamefont {S.}~\bibnamefont
  {John}},\ }\href {\doibase 10.1088/1367-2630/15/8/083033} {\bibfield
  {journal} {\bibinfo  {journal} {New Journal of Physics}\ }\textbf {\bibinfo
  {volume} {15}},\ \bibinfo {pages} {083033} (\bibinfo {year}
  {2013})}\BibitemShut {NoStop}%
\bibitem [{\citenamefont {Agarwal}\ and\ \citenamefont
  {Gupta}(1998)}]{Agarwal98}%
  \BibitemOpen
  \bibfield  {author} {\bibinfo {author} {\bibfnamefont {G.~S.}\ \bibnamefont
  {Agarwal}}\ and\ \bibinfo {author} {\bibfnamefont {S.~D.}\ \bibnamefont
  {Gupta}},\ }\href {\doibase 10.1103/PhysRevA.57.667} {\bibfield  {journal}
  {\bibinfo  {journal} {Phys. Rev. A}\ }\textbf {\bibinfo {volume} {57}},\
  \bibinfo {pages} {667} (\bibinfo {year} {1998})}\BibitemShut {NoStop}%
\bibitem [{\citenamefont {Hopmeier}\ \emph {et~al.}(1999)\citenamefont
  {Hopmeier}, \citenamefont {Guss}, \citenamefont {Deussen}, \citenamefont
  {G\"obel},\ and\ \citenamefont {Mahrt}}]{Hopmeier99}%
  \BibitemOpen
  \bibfield  {author} {\bibinfo {author} {\bibfnamefont {M.}~\bibnamefont
  {Hopmeier}}, \bibinfo {author} {\bibfnamefont {W.}~\bibnamefont {Guss}},
  \bibinfo {author} {\bibfnamefont {M.}~\bibnamefont {Deussen}}, \bibinfo
  {author} {\bibfnamefont {E.~O.}\ \bibnamefont {G\"obel}}, \ and\ \bibinfo
  {author} {\bibfnamefont {R.~F.}\ \bibnamefont {Mahrt}},\ }\href {\doibase
  10.1103/PhysRevLett.82.4118} {\bibfield  {journal} {\bibinfo  {journal}
  {Phys. Rev. Lett.}\ }\textbf {\bibinfo {volume} {82}},\ \bibinfo {pages}
  {4118} (\bibinfo {year} {1999})}\BibitemShut {NoStop}%
\bibitem [{\citenamefont {Haugland}\ \emph {et~al.}(2021)\citenamefont
  {Haugland}, \citenamefont {Sch{\"a}fer}, \citenamefont {Ronca}, \citenamefont
  {Rubio},\ and\ \citenamefont {Koch}}]{Haugland21}%
  \BibitemOpen
  \bibfield  {author} {\bibinfo {author} {\bibfnamefont {T.~S.}\ \bibnamefont
  {Haugland}}, \bibinfo {author} {\bibfnamefont {C.}~\bibnamefont
  {Sch{\"a}fer}}, \bibinfo {author} {\bibfnamefont {E.}~\bibnamefont {Ronca}},
  \bibinfo {author} {\bibfnamefont {A.}~\bibnamefont {Rubio}}, \ and\ \bibinfo
  {author} {\bibfnamefont {H.}~\bibnamefont {Koch}},\ }\href {\doibase
  10.1063/5.0039256} {\bibfield  {journal} {\bibinfo  {journal} {The Journal of
  Chemical Physics}\ }\textbf {\bibinfo {volume} {154}},\ \bibinfo {pages}
  {094113} (\bibinfo {year} {2021})}\BibitemShut {NoStop}%
\bibitem [{\citenamefont {John}\ and\ \citenamefont {Quang}(1995)}]{John95}%
  \BibitemOpen
  \bibfield  {author} {\bibinfo {author} {\bibfnamefont {S.}~\bibnamefont
  {John}}\ and\ \bibinfo {author} {\bibfnamefont {T.}~\bibnamefont {Quang}},\
  }\href {\doibase 10.1103/PhysRevA.52.4083} {\bibfield  {journal} {\bibinfo
  {journal} {Phys. Rev. A}\ }\textbf {\bibinfo {volume} {52}},\ \bibinfo
  {pages} {4083} (\bibinfo {year} {1995})}\BibitemShut {NoStop}%
\bibitem [{\citenamefont {Bay}\ \emph {et~al.}(1997{\natexlab{a}})\citenamefont
  {Bay}, \citenamefont {Lambropoulos},\ and\ \citenamefont
  {M\o{}lmer}}]{Bay97a}%
  \BibitemOpen
  \bibfield  {author} {\bibinfo {author} {\bibfnamefont {S.}~\bibnamefont
  {Bay}}, \bibinfo {author} {\bibfnamefont {P.}~\bibnamefont {Lambropoulos}}, \
  and\ \bibinfo {author} {\bibfnamefont {K.}~\bibnamefont {M\o{}lmer}},\ }\href
  {\doibase 10.1103/PhysRevLett.79.2654} {\bibfield  {journal} {\bibinfo
  {journal} {Phys. Rev. Lett.}\ }\textbf {\bibinfo {volume} {79}},\ \bibinfo
  {pages} {2654} (\bibinfo {year} {1997}{\natexlab{a}})}\BibitemShut {NoStop}%
\bibitem [{\citenamefont {Bay}\ \emph {et~al.}(1997{\natexlab{b}})\citenamefont
  {Bay}, \citenamefont {Lambropoulos},\ and\ \citenamefont
  {M\o{}lmer}}]{Bay97b}%
  \BibitemOpen
  \bibfield  {author} {\bibinfo {author} {\bibfnamefont {S.}~\bibnamefont
  {Bay}}, \bibinfo {author} {\bibfnamefont {P.}~\bibnamefont {Lambropoulos}}, \
  and\ \bibinfo {author} {\bibfnamefont {K.}~\bibnamefont {M\o{}lmer}},\ }\href
  {\doibase 10.1103/PhysRevA.55.1485} {\bibfield  {journal} {\bibinfo
  {journal} {Phys. Rev. A}\ }\textbf {\bibinfo {volume} {55}},\ \bibinfo
  {pages} {1485} (\bibinfo {year} {1997}{\natexlab{b}})}\BibitemShut {NoStop}%
\bibitem [{\citenamefont {Xie}\ \emph {et~al.}(2003)\citenamefont {Xie},
  \citenamefont {Yang}, \citenamefont {Chen},\ and\ \citenamefont
  {Zhu}}]{Xie03}%
  \BibitemOpen
  \bibfield  {author} {\bibinfo {author} {\bibfnamefont {S.-Y.}\ \bibnamefont
  {Xie}}, \bibinfo {author} {\bibfnamefont {Y.-P.}\ \bibnamefont {Yang}},
  \bibinfo {author} {\bibfnamefont {H.}~\bibnamefont {Chen}}, \ and\ \bibinfo
  {author} {\bibfnamefont {S.-Y.}\ \bibnamefont {Zhu}},\ }\href {\doibase
  10.1080/09500340308234533} {\bibfield  {journal} {\bibinfo  {journal}
  {Journal of Modern Optics}\ }\textbf {\bibinfo {volume} {50}},\ \bibinfo
  {pages} {83} (\bibinfo {year} {2003})}\BibitemShut {NoStop}%
\bibitem [{\citenamefont {Kurizki}(1990)}]{Kurizki90}%
  \BibitemOpen
  \bibfield  {author} {\bibinfo {author} {\bibfnamefont {G.}~\bibnamefont
  {Kurizki}},\ }\href {\doibase 10.1103/PhysRevA.42.2915} {\bibfield  {journal}
  {\bibinfo  {journal} {Phys. Rev. A}\ }\textbf {\bibinfo {volume} {42}},\
  \bibinfo {pages} {2915} (\bibinfo {year} {1990})}\BibitemShut {NoStop}%
\bibitem [{\citenamefont {Cortes}\ and\ \citenamefont
  {Jacob}(2017)}]{Cortes17}%
  \BibitemOpen
  \bibfield  {author} {\bibinfo {author} {\bibfnamefont {C.~L.}\ \bibnamefont
  {Cortes}}\ and\ \bibinfo {author} {\bibfnamefont {Z.}~\bibnamefont {Jacob}},\
  }\href {\doibase 10.1038/ncomms14144} {\bibfield  {journal} {\bibinfo
  {journal} {Nature Communications}\ }\textbf {\bibinfo {volume} {8}},\
  \bibinfo {pages} {14144} (\bibinfo {year} {2017})}\BibitemShut {NoStop}%
\bibitem [{\citenamefont {Behunin}\ and\ \citenamefont {Hu}(2010)}]{Behunin10}%
  \BibitemOpen
  \bibfield  {author} {\bibinfo {author} {\bibfnamefont {R.~O.}\ \bibnamefont
  {Behunin}}\ and\ \bibinfo {author} {\bibfnamefont {B.-L.}\ \bibnamefont
  {Hu}},\ }\href {\doibase 10.1103/PhysRevA.82.022507} {\bibfield  {journal}
  {\bibinfo  {journal} {Phys. Rev. A}\ }\textbf {\bibinfo {volume} {82}},\
  \bibinfo {pages} {022507} (\bibinfo {year} {2010})}\BibitemShut {NoStop}%
\bibitem [{\citenamefont {Jones}\ \emph {et~al.}(2018)\citenamefont {Jones},
  \citenamefont {Needham}, \citenamefont {Lesanovsky}, \citenamefont
  {Intravaia},\ and\ \citenamefont {Olmos}}]{Jones18}%
  \BibitemOpen
  \bibfield  {author} {\bibinfo {author} {\bibfnamefont {R.}~\bibnamefont
  {Jones}}, \bibinfo {author} {\bibfnamefont {J.~A.}\ \bibnamefont {Needham}},
  \bibinfo {author} {\bibfnamefont {I.}~\bibnamefont {Lesanovsky}}, \bibinfo
  {author} {\bibfnamefont {F.}~\bibnamefont {Intravaia}}, \ and\ \bibinfo
  {author} {\bibfnamefont {B.}~\bibnamefont {Olmos}},\ }\href {\doibase
  10.1103/PhysRevA.97.053841} {\bibfield  {journal} {\bibinfo  {journal} {Phys.
  Rev. A}\ }\textbf {\bibinfo {volume} {97}},\ \bibinfo {pages} {053841}
  (\bibinfo {year} {2018})}\BibitemShut {NoStop}%
\bibitem [{\citenamefont {Sinha}\ \emph {et~al.}(2018)\citenamefont {Sinha},
  \citenamefont {Venkatesh},\ and\ \citenamefont {Meystre}}]{CollCP18}%
  \BibitemOpen
  \bibfield  {author} {\bibinfo {author} {\bibfnamefont {K.}~\bibnamefont
  {Sinha}}, \bibinfo {author} {\bibfnamefont {B.~P.}\ \bibnamefont
  {Venkatesh}}, \ and\ \bibinfo {author} {\bibfnamefont {P.}~\bibnamefont
  {Meystre}},\ }\href {\doibase 10.1103/PhysRevLett.121.183605} {\bibfield
  {journal} {\bibinfo  {journal} {Phys. Rev. Lett.}\ }\textbf {\bibinfo
  {volume} {121}},\ \bibinfo {pages} {183605} (\bibinfo {year}
  {2018})}\BibitemShut {NoStop}%
\bibitem [{\citenamefont {Yang}\ \emph {et~al.}(2020)\citenamefont {Yang},
  \citenamefont {Khandekar}, \citenamefont {Li},\ and\ \citenamefont
  {Jacob}}]{Yang2020}%
  \BibitemOpen
  \bibfield  {author} {\bibinfo {author} {\bibfnamefont {L.-P.}\ \bibnamefont
  {Yang}}, \bibinfo {author} {\bibfnamefont {C.}~\bibnamefont {Khandekar}},
  \bibinfo {author} {\bibfnamefont {T.}~\bibnamefont {Li}}, \ and\ \bibinfo
  {author} {\bibfnamefont {Z.}~\bibnamefont {Jacob}},\ }\href {\doibase
  10.1088/1367-2630/ab6f92} {\bibfield  {journal} {\bibinfo  {journal} {New
  Journal of Physics}\ }\textbf {\bibinfo {volume} {22}},\ \bibinfo {pages}
  {023037} (\bibinfo {year} {2020})}\BibitemShut {NoStop}%
\bibitem [{\citenamefont {Varada}\ and\ \citenamefont
  {Agarwal}(1992)}]{Varada92}%
  \BibitemOpen
  \bibfield  {author} {\bibinfo {author} {\bibfnamefont {G.~V.}\ \bibnamefont
  {Varada}}\ and\ \bibinfo {author} {\bibfnamefont {G.~S.}\ \bibnamefont
  {Agarwal}},\ }\href {\doibase 10.1103/PhysRevA.45.6721} {\bibfield  {journal}
  {\bibinfo  {journal} {Phys. Rev. A}\ }\textbf {\bibinfo {volume} {45}},\
  \bibinfo {pages} {6721} (\bibinfo {year} {1992})}\BibitemShut {NoStop}%
\bibitem [{\citenamefont {de~L\'es\'eleuc}\ \emph {et~al.}(2017)\citenamefont
  {de~L\'es\'eleuc}, \citenamefont {Barredo}, \citenamefont {Lienhard},
  \citenamefont {Browaeys},\ and\ \citenamefont {Lahaye}}]{deLeseleuc17}%
  \BibitemOpen
  \bibfield  {author} {\bibinfo {author} {\bibfnamefont {S.}~\bibnamefont
  {de~L\'es\'eleuc}}, \bibinfo {author} {\bibfnamefont {D.}~\bibnamefont
  {Barredo}}, \bibinfo {author} {\bibfnamefont {V.}~\bibnamefont {Lienhard}},
  \bibinfo {author} {\bibfnamefont {A.}~\bibnamefont {Browaeys}}, \ and\
  \bibinfo {author} {\bibfnamefont {T.}~\bibnamefont {Lahaye}},\ }\href
  {\doibase 10.1103/PhysRevLett.119.053202} {\bibfield  {journal} {\bibinfo
  {journal} {Phys. Rev. Lett.}\ }\textbf {\bibinfo {volume} {119}},\ \bibinfo
  {pages} {053202} (\bibinfo {year} {2017})}\BibitemShut {NoStop}%
\bibitem [{\citenamefont {Pedrotti}\ \emph {et~al.}(2006)\citenamefont
  {Pedrotti}, \citenamefont {Pedrotti},\ and\ \citenamefont
  {Pedrotti}}]{Pedrotti}%
  \BibitemOpen
  \bibfield  {author} {\bibinfo {author} {\bibfnamefont {F.}~\bibnamefont
  {Pedrotti}}, \bibinfo {author} {\bibfnamefont {L.~S.}\ \bibnamefont
  {Pedrotti}}, \ and\ \bibinfo {author} {\bibfnamefont {L.~M.}\ \bibnamefont
  {Pedrotti}},\ }\href@noop {} {\emph {\bibinfo {title} {Introduction to
  Optics}}}\ (\bibinfo  {publisher} {Pearson Education},\ \bibinfo {year}
  {2006})\ p.\ \bibinfo {pages} {448}\BibitemShut {NoStop}%
\bibitem [{\citenamefont {Gruner}\ and\ \citenamefont
  {Welsch}(1996)}]{Gruner96}%
  \BibitemOpen
  \bibfield  {author} {\bibinfo {author} {\bibfnamefont {T.}~\bibnamefont
  {Gruner}}\ and\ \bibinfo {author} {\bibfnamefont {D.-G.}\ \bibnamefont
  {Welsch}},\ }\href {\doibase 10.1103/PhysRevA.53.1818} {\bibfield  {journal}
  {\bibinfo  {journal} {Phys. Rev. A}\ }\textbf {\bibinfo {volume} {53}},\
  \bibinfo {pages} {1818} (\bibinfo {year} {1996})}\BibitemShut {NoStop}%
\bibitem [{\citenamefont {Buhmann}\ and\ \citenamefont
  {Welsch}(2007)}]{BuhmannRev}%
  \BibitemOpen
  \bibfield  {author} {\bibinfo {author} {\bibfnamefont {S.~Y.}\ \bibnamefont
  {Buhmann}}\ and\ \bibinfo {author} {\bibfnamefont {D.-G.}\ \bibnamefont
  {Welsch}},\ }\href {\doibase
  https://doi.org/10.1016/j.pquantelec.2007.03.001} {\bibfield  {journal}
  {\bibinfo  {journal} {Progress in Quantum Electronics}\ }\textbf {\bibinfo
  {volume} {31}},\ \bibinfo {pages} {51} (\bibinfo {year} {2007})}\BibitemShut
  {NoStop}%
\bibitem [{\citenamefont {Buhmann}(2012{\natexlab{a}})}]{Buhmann1}%
  \BibitemOpen
  \bibfield  {author} {\bibinfo {author} {\bibfnamefont {S.~Y.}\ \bibnamefont
  {Buhmann}},\ }\href@noop {} {\emph {\bibinfo {title} {Dispersion Forces I}}}\
  (\bibinfo  {publisher} {Springer-Verlag, Berlin, Heidelberg},\ \bibinfo
  {year} {2012})\BibitemShut {NoStop}%
\bibitem [{\citenamefont {Buhmann}(2012{\natexlab{b}})}]{Buhmann2}%
  \BibitemOpen
  \bibfield  {author} {\bibinfo {author} {\bibfnamefont {S.~Y.}\ \bibnamefont
  {Buhmann}},\ }\href@noop {} {\emph {\bibinfo {title} {Dispersion Forces
  II}}}\ (\bibinfo  {publisher} {Springer-Verlag, Berlin, Heidelberg},\
  \bibinfo {year} {2012})\BibitemShut {NoStop}%
\bibitem [{\citenamefont {Breuer}\ and\ \citenamefont
  {Petruccione}(2002)}]{BPbook}%
  \BibitemOpen
  \bibfield  {author} {\bibinfo {author} {\bibfnamefont {H.-P.}\ \bibnamefont
  {Breuer}}\ and\ \bibinfo {author} {\bibfnamefont {F.}~\bibnamefont
  {Petruccione}},\ }\href@noop {} {\emph {\bibinfo {title} {Theory of open
  quantum systems}}}\ (\bibinfo  {publisher} {Oxford University Press},\
  \bibinfo {address} {New York},\ \bibinfo {year} {2002})\BibitemShut {NoStop}%
\bibitem [{\citenamefont {Chang}\ \emph {et~al.}(2014)\citenamefont {Chang},
  \citenamefont {Sinha}, \citenamefont {Taylor},\ and\ \citenamefont
  {Kimble}}]{Chang14}%
  \BibitemOpen
  \bibfield  {author} {\bibinfo {author} {\bibfnamefont {D.~E.}\ \bibnamefont
  {Chang}}, \bibinfo {author} {\bibfnamefont {K.}~\bibnamefont {Sinha}},
  \bibinfo {author} {\bibfnamefont {J.~M.}\ \bibnamefont {Taylor}}, \ and\
  \bibinfo {author} {\bibfnamefont {H.~J.}\ \bibnamefont {Kimble}},\ }\href
  {\doibase 10.1038/ncomms5343} {\bibfield  {journal} {\bibinfo  {journal}
  {Nature Communications}\ }\textbf {\bibinfo {volume} {5}},\ \bibinfo {pages}
  {4343} (\bibinfo {year} {2014})}\BibitemShut {NoStop}%
\bibitem [{\citenamefont {Novotny}\ and\ \citenamefont
  {Hecht}(2012)}]{Novotnybook}%
  \BibitemOpen
  \bibfield  {author} {\bibinfo {author} {\bibfnamefont {L.}~\bibnamefont
  {Novotny}}\ and\ \bibinfo {author} {\bibfnamefont {B.}~\bibnamefont
  {Hecht}},\ }\href@noop {} {\emph {\bibinfo {title} {Principles of
  nano-optics}}}\ (\bibinfo  {publisher} {Cambridge University Press},\
  \bibinfo {address} {Cambridge},\ \bibinfo {year} {2012})\BibitemShut
  {NoStop}%
\bibitem [{\citenamefont {Eldredge}\ \emph {et~al.}(2016)\citenamefont
  {Eldredge}, \citenamefont {Solano}, \citenamefont {Chang},\ and\
  \citenamefont {Gorshkov}}]{Eldredge}%
  \BibitemOpen
  \bibfield  {author} {\bibinfo {author} {\bibfnamefont {Z.}~\bibnamefont
  {Eldredge}}, \bibinfo {author} {\bibfnamefont {P.}~\bibnamefont {Solano}},
  \bibinfo {author} {\bibfnamefont {D.}~\bibnamefont {Chang}}, \ and\ \bibinfo
  {author} {\bibfnamefont {A.~V.}\ \bibnamefont {Gorshkov}},\ }\href {\doibase
  10.1103/PhysRevA.94.053855} {\bibfield  {journal} {\bibinfo  {journal} {Phys.
  Rev. A}\ }\textbf {\bibinfo {volume} {94}},\ \bibinfo {pages} {053855}
  (\bibinfo {year} {2016})}\BibitemShut {NoStop}%
\bibitem [{\citenamefont {Chang}\ \emph {et~al.}(2013)\citenamefont {Chang},
  \citenamefont {Cirac},\ and\ \citenamefont {Kimble}}]{Chang2013}%
  \BibitemOpen
  \bibfield  {author} {\bibinfo {author} {\bibfnamefont {D.~E.}\ \bibnamefont
  {Chang}}, \bibinfo {author} {\bibfnamefont {J.~I.}\ \bibnamefont {Cirac}}, \
  and\ \bibinfo {author} {\bibfnamefont {H.~J.}\ \bibnamefont {Kimble}},\
  }\href {\doibase 10.1103/PhysRevLett.110.113606} {\bibfield  {journal}
  {\bibinfo  {journal} {Phys. Rev. Lett.}\ }\textbf {\bibinfo {volume} {110}},\
  \bibinfo {pages} {113606} (\bibinfo {year} {2013})}\BibitemShut {NoStop}%
\bibitem [{\citenamefont {Steck}()}]{CesiumData}%
  \BibitemOpen
  \bibfield  {author} {\bibinfo {author} {\bibfnamefont {D.~A.}\ \bibnamefont
  {Steck}},\ }\href@noop {} {}\bibinfo {howpublished} {"Cesium D Line Data,"
  available online at \url{http://steck.us/alkalidata}},\ \bibinfo {note}
  {(revision 2.2.1, 21 November 2019)}\BibitemShut {NoStop}%
\bibitem [{\citenamefont {Hohenester}(2019)}]{Hohenesterbook}%
  \BibitemOpen
  \bibfield  {author} {\bibinfo {author} {\bibfnamefont {U.}~\bibnamefont
  {Hohenester}},\ }\href@noop {} {\emph {\bibinfo {title} {Nano and Quantum
  Optics: An Introduction to Basic Principles and Theory}}}\ (\bibinfo
  {publisher} {Springer},\ \bibinfo {address} {Switzerland},\ \bibinfo {year}
  {2019})\BibitemShut {NoStop}%
\end{thebibliography}%

\end{document}